\documentclass[useAMS,usenatbib]{mn2e}
\usepackage{graphicx}
\usepackage[hang]{subfigure}
\usepackage{ amssymb }
\usepackage{fmtcount}
\usepackage{ textcomp }
\usepackage{ gensymb }

\makeatletter
\DeclareMathSizes{12}{12}{7}{7}
\makeatletter

\def\Msun{\hbox{$\rm\thinspace M_{\odot}$}}
\def\asec{$^{\prime\prime}$}
\def\MUV{$M_{1500}$}

\title[LBG size evolution]{Non-parametric analysis of the rest-frame UV sizes and morphological disturbance amongst L$_{*}$ galaxies at  $4<z<8$}
\author[E. Curtis-Lake et al.]{E. Curtis-Lake$^{1,2}$\thanks{Email: curtis@iap.fr},
R.J. McLure$^{1}$, J.S. Dunlop$^{1}$, A.B. Rogers$^{1}$, T. Targett$^{1}$, 
\newauthor A. Dekel$^{3}$, R.S. Ellis$^{4}$, S.M. Faber$^{5}$,
 H.C. Ferguson$^{6}$, N.A. Grogin$^{6}$, 
\newauthor D.D. Kocevski$^{5}$, A.M. Koekemoer$^{6}$, K. Lai$^{5}$, 
E. M\'armol-Queralt\'o$^{1}$,
\newauthor  B.E. Robertson$^{7}$\\
$^{1}$ SUPA\thanks{Scottish Universities Physics Alliance},
Institute for Astronomy, University of Edinburgh, 
Royal Observatory, Edinburgh EH9 3HJ\\
$^{2}$ Sorbonne Universit\'es, UPMC-CNRS, UMR7095, Institut d'Astrophysique de Paris, F-75014, Paris, France\\
$^{3}$ Center for Astrophysics and Planetary Science, Racah Institute of Physics, The Hebrew University, Jerusalem 91904 Israel\\
$^{4}$ Department of Astronomy, California Institute of Technology, Pasadena, CA 91125, USA\\
$^{5}$ University of California Observatories/Lick Observatory, University of California, Santa Cruz, CA 95064, USA\\
$^{6}$ Space Telescope Science Institute, 3700 San Martin Drive, Baltimore, MD 21218, USA\\
$^{7}$ Department of Astronomy, University of Arizona, 933 North Cerry Avenue, Tucson, AZ 85721, USA}

\begin{document}

\date{}

\pagerange{\pageref{firstpage}--\pageref{lastpage}} \pubyear{2002}
\maketitle

\label{firstpage}
\def\Msun{\hbox{$\rm\thinspace M_{\odot}$}}

\begin{abstract}
We present the results of a study investigating the sizes and morphologies of redshift $4~<~z~<~8$ galaxies in the CANDELS (Cosmic Assembly Near-Infrared Deep Extragalactic Legacy Survey) GOODS-S (Great Observatories Origins Deep Survey southern field), HUDF (\textit{Hubble} Ultra-Deep Field) and HUDF parallel fields. Based on non-parametric measurements and incorporating a careful treatment of measurement biases, we quantify the {\it typical} size of galaxies at each redshift as the \textit{peak} of the log-normal size distribution, rather than the arithmetic mean size. Parameterizing the evolution of galaxy half-light radius as $r_{50}\propto(1+z)^n$, we find $n=-0.20\pm0.26$ at bright UV-luminosities ($0.3\textrm{L}_{*(z=3)}<\textrm{L}<\textrm{L}_*$) and $n=-0.47\pm0.62$ at faint luminosities ($0.12\textrm{L}_*<\textrm{L}<0.3\textrm{L}_*$). Furthermore, simulations based on artificially redshifting our $z\sim4$ galaxy sample show that we cannot reject the null hypothesis of no size evolution.  We show that this result is caused by a combination of the size-dependent completeness of high-redshift galaxy samples and the underestimation of the sizes of the largest galaxies at a given epoch. To explore the evolution of galaxy morphology we first compare asymmetry measurements to those from a large sample of simulated single S\'ersic profiles, in order to robustly categorise galaxies as either `smooth' or `disturbed'. Comparing the disturbed fraction amongst bright (\MUV$~\leq -20$) galaxies at each redshift to that obtained by artificially redshifting our $z\sim 4$ galaxy sample, while carefully matching the size and UV-luminosity distributions, we find no clear evidence for evolution in galaxy morphology over the redshift interval $4<z<8$. Therefore, based on our results, a bright (\MUV$~\leq -20$) galaxy at $z\sim6$ is no more likely to be measured as `disturbed' than a comparable galaxy at $z\sim 4$, given the current observational constraints.
\end{abstract}


\newpage

\section{Introduction}

The best constraints currently available for discerning how the first galaxies formed are derived from ultra-violet (UV) selected samples.  These are star-forming galaxies by definition and analysing their structure in the rest frame UV can provide important information about the physical mechanisms responsible for that star-formation.

The high-redshift galaxy luminosity function (e.g. \citealt{Bouwens2007,McLure2009,McLure2013,Schenker2013,Lorenzoni2012}) and spectral energy distribution (SED) fitting (e.g. \citealt{Jiang2013a,Curtis-Lake2013,Stark2009,Stark2013}) can tell us about the evolving abundances and stellar populations of these high-redshift galaxies, while measuring their sizes and morphologies (e.g. \citealt{Ferguson2004,Bouwens2004,Hathi2008a}; \citealt*{Conselice2009}; \citealt{Oesch2010a,Jiang2013a,Ono2013}) provides us with complementary information of how they grow and evolve.

A framework was laid out for understanding how the sizes of disc galaxies can be related to the evolution of their parent haloes by \cite*{Mo1998}, according to the disc formation model of \cite{FallS.M.1980}.
The Mo et al. (1998) formalism assumes that disc masses and angular momenta are fixed fractions of those of their parent dark matter haloes.  This in turn predicts that galaxy sizes should evolve $\propto H(z)^{-1}$ ($\propto (1+z)^{-3/2}$) at constant halo circular velocity, or $\propto H(z)^{-2/3}$ ($\propto (1+z)^{-1}$) at constant halo mass.   This relies on many assumptions, including a redshift invariant dark matter halo profile.

However, at high redshift we are observationally forced to study Lyman break galaxy (LBG) sizes from the rest-frame UV, at approximately constant UV luminosity.  This selection does not necessarily follow constant halo velocity or halo mass,  complicating the interpretation of the inferred evolution. The exponent of the $(1+z)^n$ relation fitted to the data therefore only reveals whether the UV luminosity most closely traces the halo velocity or halo mass \textit{if} all the other assumptions hold. 
Early studies disagreed on the size evolution with the study of \cite{Bouwens2004a} finding evolution close to that expected for constant halo mass evolution and \cite{Ferguson2004} finding evolution described best by selection at constant circular velocity.  More recent studies, however, have measured the size evolution to be somewhere between the two scenarios ($\propto(1+z)^{-1.12\pm0.17}$ for bright galaxies and  $\propto(1+z)^{-1.32\pm0.52}$ for fainter galaxies in \citealt{Oesch2010a}, and $\propto(1+z)^{-1.30\pm0.13}$ in \citealt{Ono2013}).

Moving beyond measurements of galaxy size evolution, the evolution of galaxy morphology is clearly of interest.
Without analysing galaxy morphologies we cannot address some important questions.  Are major mergers important at high redshifts? Is the star formation evenly distributed or is it occurring in distinct clumps as shown in lower redshift clump-cluster galaxies \citep{Elmegreen2005}? 
How would these factors affect the inferred size evolution when measured from the rest-frame UV?

Some studies have already attempted to categorise the morphologies of LBG samples using CAS/Gini/M20 measurements as well as visual inspection (e.g. Conselice et al. 2009; \citealt{Jiang2013a}).
\cite{Jiang2013a} found the merger rate at the bright end of a sample of $z\sim6$ LBGs to be as high as $ \sim 48\%$.  Although they investigated applying Gini/CAS/M20 measurements to their sample, they found that the most reliable way to distinguish interacting galaxies was by visual inspection.  They concluded that the small object sizes meant that the interacting systems were not easily differentiated using non-parametric measurements alone.  Conselice et al. (2009) investigated  the morphologies of a sample of $4<z<6$ LBGs finding that $\sim30\%$ of the galaxies showed distorted and asymmetric structures, and found marginal evidence that the distorted galaxies had higher star formation rates (SFRs) than their smooth counterparts.

The aim of this study is to investigate the fraction of disturbed galaxy morphologies in the LBG population, testing for any evidence of its evolution and examining the links with the observed size evolution.  We use consistent rest-frame wavelengths for the size measurements, and measure the sizes and morphologies non-parametrically incorporating a proper treatment of the biases inherent in the measurements.

Sizes are measured with a simple, non-parametric curve of growth (CoG) and a simple diagnostic (the non-parametric asymmetry measurement, \citealt*{Conselice2000}) is used to determine whether a galaxy can be distinguished from a smooth, symmetric profile.  The method we employ robustly takes account of surface brightness and resolution effects on an image-by-image basis. 

This morphological diagnostic is image-dependent and so careful analysis is required when investigating any evidence for evolution.  This is done by comparing measurements to those derived from an artificially redshifted (AR) $z\sim4$ galaxy sample, allowing us to investigate any evidence for evolution in the fraction of disturbed galaxies all the way up to $z\sim8$.  This AR sample is also used as a test of whether we can reject the null hypothesis for size evolution across the redshift range studied, given our sampling of the galaxy population.  Testing against the artificially redshifted $z\sim4$ sample provides a consistent method for quantifying the significance of any measured evolution

The structure of the paper is as follows. In Section 2 the data and sample selection is described.  The non-parametric size and asymmetry measurements are described in Section 3. In Section 4 the simulations used to assess different size measurement techniques, distinguish galaxies that are not consistent with smooth, symmetric profiles, as well as the AR samples are described.  The results are presented in Section 5 and discussed in Section 6.  Finally, the conclusions are presented in Section 7.  Throughout this paper standard cosmology is assumed, with $H_0 = 70$km s$^{-1}$ Mpc$^{-3}$, $\Omega_m=0.3$ and $\Omega_{\Lambda}=0.7$.

\section{Data}

\subsection{Imaging Data}

The samples were selected from regions with deep near-infrared (NIR) Wide-Field Camera 3 (WFC3) and optical Advanced Camera for Surveys (ACS) imaging within three main surveys: Cosmic Assembly Near-infrared Deep Extragalactic Legacy Survey (CANDELS; \citealt{Grogin2011,Koekemoer2011}), Ultra-Deep Field 2012 (HUDF12; \citealt{Ellis2013,Koekemoer2013}) and the \textit{Hubble} Ultra-Deep Field 2009 (HUDF09; \citealt{Bouwens2011}).  From the CANDELS survey we used the data covering the Great Observatories Origins Deep Survey southern field (GOODS-S) to provide measurements of brighter objects.  For measurements of fainter objects, samples were taken from the HUDF and its two parallel fields.  A summary of the depths and filters available in each of these fields can be found in \cite{McLure2013}. All analysis was performed on 60mas pixel-scale mosaics.

\subsubsection{GOODS-S}

To provide coverage at the bright end of the high-redshift ($z\geq4$) luminosity function, the publicly available WFC3/IR imaging of GOODS-S with the F105W, F125W and F160W filters (hereinafter referred to as $Y_{105}$, $J_{125}$ and $H_{160}$ respectively) provided by the CANDELS survey \citep{Grogin2011,Koekemoer2011} was combined with the v2.0 reduction of the publicly available ACS data  \citep{Giavalisco2004} in the optical filters; F435W, F606W, F775W, F850LP (hereinafter referred to as $B_{435}$, $V_{606}$, $i_{775}$and $z_{850}$ respectively).
This study makes use of the deep, wide and ERS (Early Release Science) regions of GOODS-S WFC3/IR imaging.  In the ERS, deep F098W ($Y_{098}$) imaging is available \citep{Windhorst2011}, rather than $Y_{105}$ which is available over the rest of the GOODS-S CANDELS imaging.  

\subsubsection{HUDF12}

To add faint galaxies to the sample, LBGs are selected from the HUDF and parallel fields.  The most recent coverage of the HUDF in the NIR was provided by the HUDF12 survey \citep{Ellis2013,Koekemoer2013} and was utilised here.  This includes deeper coverage in the $Y_{105}$ and $H_{160}$ filters, as well as a new deep F140W ($J_{140}$) image. ACS imaging from the \cite{Beckwith2006} HUDF ACS programme was used to provide optical coverage in the $B_{435}$, $V_{606}$, $i_{775}$ and $z_{850}$ filters.

\subsubsection{HUDF09 parallel fields}

Galaxies were also selected from the two deep HUDF parallel fields.  A new reduction of the NIR data taken as part of the HUDF09 campaign \citep{Bouwens2011} was used for both parallel fields \citep{Koekemoer2013}.  In the first parallel field (P1, field centre: $03^{h}33^{m}03.60^{s}$, $-27\degree51^{\prime}01.80^{\prime\prime}$), we used publicly available mosaics of the ACS data originally obtained as part of the HUDF05 campaign (GO-10632, P.I. Stiavelli), while a new reduction of the same data was used in the second parallel field (P2, field centre: $03^{h}33^{m}07.75^{s}$, $-27\degree51^{\prime}47.00^{\prime\prime}$).  Only one of the fields has $B_{435}$-band coverage (P2) but this imaging is $\sim1$ mag shallower than $V_{606}$ (see \cite{Bouwens2011}, Table 1) and so is not used in this paper.

\subsection{Point Spread Functions}

The point spread functions (PSFs) used throughout this work are made from stars cutout from the images themselves.  For the CANDELS imaging in the deep, wide and ERS fields, the PSFs were made from median stacking bright, unsaturated, uncontaminated stars within the field.  Before stacking the stars were centered using the {\scshape idl} procedure GCNTRD, background subtracted and flux normalised. In the case of the HUDF and parallel fields a single bright, unsaturated, uncontaminated star was used.  The PSFs have measured half-light radii (within a 20 pixel radius total flux aperture) of $\sim 0.045$\arcsec$-0.055$\arcsec in the ACS images ($f_{606}$, $f_{775}$ , and $f_{850}$) and $\sim 0.065$\arcsec$-0.080$\arcsec in the WFC3 images ( $f_{105}$, $f_{125}$, $f_{140}$ and  $f_{160}$). The measured $\sigma$ values from fitting a Gaussian profile to the central regions of the PSFs are $\sim 0.065$\arcsec in the ACS images ($f_{606}$, $f_{775}$ , and $f_{850}$) and $\sim 0.095$\arcsec$-0.1$\arcsec in the WFC3 images ( $f_{105}$, $f_{125}$, $f_{140}$ and  $f_{160}$) 

\subsection{Selection}
\label{section:Selection}

Photometric redshifts were used to select galaxies with $z_{phot} > 3.5$ from the fields summarised above.  The catalogues used to measure these photometric redshifts were produced using {\scshape SExtractor} as described below.

Separate catalogues were produced by using each image in turn as a detection image and using {\scshape SExtractor} in dual image mode.  At least one filter was required to be bluewards of the detection image so that the Lyman break (the strongest spectral feature for photometric redshift identification at $z\gtrsim3.5$) was always bracketed by two filters.  The shortest wavelength detection images were therefore $V_{606}$ in CANDELS deep, wide, ERS and the HUDF field.  In the HUDF parallels, however, the shortest wavelength detection image was $i_{775}$.

Aperture photometry was performed using 0.3\asec (5 pixel) diameter apertures in the $B_{435}$, $V_{606}$ and $i_{775}$ images, a 0.42\asec (7 pixel) diameter aperture in the $z_{850}$ image and 0.48\asec (8 pixel) radius apertures in the $Y_{105}$, $J_{125}$, $J_{140}$ and $H_{160}$ images.  Apertures were chosen to enclose at least 70\% point source flux and all photometry was corrected to total using point source aperture corrections.

Photometric errors were estimated from local image depth measurements.  The local depths were estimated from the width of the distribution of aperture fluxes from multiple apertures, with the same radius as the measurement apertures, placed in empty regions of a 60\asec~{x}~60\asec~ box centered on the object of interest.

A master catalogue of unique sources was then created containing all the objects present at $5\sigma$ in at least one image but fainter or not present in the shorter wavelength images.  For example, to make the HUDF unique source catalogue we started with objects present in the $V_{606}$ image at $5\sigma$.  The only stipulation required for objects detected in this lowest wavelength detection filter was that the $B_{435}$ detection must be fainter (as would be required by the presence of the Lyman break between the two filters).  Then objects at $5\sigma$ in the $i_{775}$ image that were not present at $5\sigma$ in the $V_{606}$ image (or $3\sigma$ in the $B_{435}$ image) are added, then all objects with $5\sigma$ in $z_{850}$, $<5\sigma$ in $i_{775}$ and $<3\sigma$ in $V_{606}$, $B_{435}$ are added etc.  Written more generally, the flux in the detection image was required to be $>5\sigma$ and the flux in the filter directly blue-wards to be $<5\sigma$, and any bluer filters to have $<3\sigma$ detections ($\sigma$ is determined from the local depth estimates).   

Photometric redshifts were then measured for all objects in the master catalogue using the Le Phare photometric redshift code \citep{Ilbert2009}.  The \cite{Ilbert2009} template set was used, which was originally used to derive photometric redshifts in the Cosmological Evolution Survey (\citealt{Scoville2007}).  This template set consists of the three SEDs of elliptical galaxies and 6 of spiral galaxies (S0, Sa, Sb, Sc, Sd, Sdm) produced by \cite{Polletta2007} and 12 additional starburst templates produced using BC03 with ages ranging from 3 to 0.03 Gyr.  Intergalactic (IGM) absorption is applied using the \cite{Madau1995} prescription and dust attenuation is included in the fitting using the \cite{Calzetti2000} dust curve with a range of extinction values, $0<E(B-V)< 1.5$.  Any objects with a high-redshift primary solution ($z_{phot,best}>3.5$) with $\chi^2_{best}<20$, and $\Delta\chi^2>2$ between the high-redshift solution and any secondary low-redshift solution, were selected.  

Stellar contaminants were identified using both the SED and half-light radius information.  Each object SED was fit using a set of L, M and T dwarf star reference spectra from the SpecX\footnote{http://pono.ucsd.edu/~adam/browndwarfs/spexprism/} library.  Objects were rejected if the best-fitting stellar template $\chi^2$ value is statistically acceptable and the size of the object is similar in size to that measured from the image PSF.  To be precise, the half-light radius must be within one pixel of the measured PSF half-light radius for the object to be rejected.  A total of 123 objects were rejected as stellar contaminants ($\sim3\%$ of the sample).  53 objects with statistically acceptable fits with sizes within 1.5 pixels of the PSF half-light radius were flagged.

At this stage visual inspection was performed on the whole sample to reject artefacts, objects with photometry contaminated by near-by low-redshift galaxies, or obvious low redshift interlopers.  The objects were then sorted into two categories: firm high-redshift candidates (flag 1) and possible high-redshift candidates (flag 2). Objects flagged as having good stellar fits were also given a flag value of 2 (see above).  Objects entering into the latter category are either very faint or have possible low-redshift solutions with $2 <\Delta\chi^2 \lesssim 4$.  Those objects with $\Delta\chi^2\lesssim4$ tend to be quite red and although this sample is likely to have a larger fraction of low-redshift interlopers, they are included here to avoid excluding reddened high-redshift galaxies from the sample.  These objects are more prevalent in the lower-redshift samples ($z_{phot} < 6$).  The sample is split in this way between firm and insecure candidates so that the effect of possible interlopers on the main results can be tested.  The final sample numbers in each field are presented in Table \ref{table:selection}. 

\begin{table}
  \centering
  \caption{Number of galaxies selected in each field and redshift bin (width of each bin is $\Delta{z}=1$). The first row for each field gives the number of galaxies selected in each redshift bin, while the second row (in bold) gives the number of objects that pass the flux cuts imposed for robust size measurements (see Section \ref{section:sizeBias}).}
  \begin{tabular}{@{}lr|ccccc@{}}
  \hline
  \hline
         & Redshift bins:& 4    & 5   & 6   & 7  & 8 \\
  \hline
  Fields:&  GOODS-S deep & 1255 & 421 & 164 & 43 & 30 \\
  
   	     &     			 & 255  & 143 & 96  & 18 & 7  \\
         &\\              
         &           ERS & 399  & 123 & 29  & 41 & 4  \\
         &   	         & 108  & 27  & 8   & 9  & 0  \\
         &\\  
         &          wide & 328  & 137 & 37  & 6  & 11 \\
  		 &		         & 71  & 36  & 10  & 1  & 0  \\
         &\\  
         &          HUDF & 229  & 106 & 72  & 23 & 10 \\
         &   		     & 108   & 67  & 53  & 16 & 9  \\
         &\\  
         &       HUDF-p1 & 13   & 54  & 28  & 11 & 8  \\
         & 		         & 7    &  31  & 11   & 8  & 4  \\
         &\\  
         &            p2 & 46   & 73  & 23  & 6  & 8  \\
         &      	     & 19   & 32  & 10  & 3  & 4 \\
         &\\
     \hline
 Legend: &  \multicolumn{6}{c}{\small{Objects passing initial selection}}\\
                  &  \multicolumn{6}{c}{\small{Objects with size measurements}}\\
  
  \hline
  \hline
\end{tabular}
\label{table:selection}
\end{table}

\section{Size and asymmetry measurements}

\subsection{Size estimates}
\label{section:sizeMeasurement}

The circularised half-light radii of the selected objects are measured from their CoG within the image closest to a rest-frame wavelength of $\lambda_{rest}=1500$\AA.  First {\scshape SExtractor} \citep{BertinE.1996} is used to produce an object mask using the segmentation map, and the image is subsampled to 1/5th of the original image scale; i.e. 0.012\asec/pixel.  
Aperture photometry is then used to measure the increase in enclosed flux as a function of radius, centering the apertures on the brightest pixel in the sub-sampled image.  

The uncertainty in half-light radii measurements is driven by two factors, the background and total flux measurements.  A large total flux aperture increases the errors in the size measurements, yet a small total flux aperture will systematically underestimate the sizes of large galaxies (see sections \ref{section:measurementCalibration} and \ref{section:sizeBias}).  Throughout the paper, a total flux aperture with a radius of 10 pixels is used to derive the size estimates for main results, but a 15 pixel radius aperture is used to test whether any of the results are strongly biased by this decision.  The measured total fluxes that are reported are derived from the 15 pixel aperture, thereby de-coupling the size and flux measurements (more details given in Sections \ref{section:measurementCalibration} and \ref{section:sizeBias}).

Although an initial modal background value is first subtracted from the images, significant background structure in the images means that a secondary background estimation is performed by requiring that the CoG is flat between two radii close to the source (the inner radius of the background annulus is 10 (15) pixels and the outer radius is 25 (30) pixels).

The sizes are PSF corrected and the fluxes aperture corrected to total using simulated single S\'ersic profiles as described in Section \ref{section:measurementCalibration}.

\subsection{Asymmetry measurements}
\label{section:asymmetryMeasurements}

Asymmetry measurements are performed according to the prescription of Conselice et al. (2000).  Essentially, the object is rotated by 180\textdegree~and subtracted from the original image.  The asymmetry is a sum of the residuals within a given radius, scaled according to the profile flux.  The centre of rotation is determined as the point at which the asymmetry is minimised and is found to 1/5th pixel precision according to the method laid out in Conselice et al. (2000).

To account for the noise in the asymmetry measurements produced by the background, the background asymmetry is calculated in blank regions of the measurement images.  The background value is then subtracted from the asymmetry value.  In practice, the background asymmetry is measured within a fixed-size radius across the whole image and then scaled according to the size of the object.  This calculation is summarised in the following equation:

\begin{equation}
A=min\left(\frac { \sum { \left| I-{ I }_{ 180 } \right|  }  }{ \sum { \left| I \right|  }  }\right) - min\left(\frac { \sum { \left| B-{ B }_{ 180 } \right|  }  }{ \sum { \left| I \right|  }  } \right)
\end{equation}

where $I$ denotes the original image pixels, $I_{180}$ are the pixels of the image rotated about its centre by 180\textdegree, $B$ are background pixels taken from a blank part of the image and $B_{180}$ are the rotated background pixels.

For asymmetry measurements to be useful, they must be measured within a radius associated with the physical scale of the object Conselice et al. (2000).  For the physical scale we use the radius enclosing 70\% of the object's flux ($r_{70}$) as measured within the 10 pixel radius aperture.  This was chosen as opposed to the Petrosian radius (used in Conselice et al. 2009), because it provides a higher signal-to-noise (S/N) measurement.

The choice of measuring the asymmetry within $r_{70}$ restricts the analysis to asymmetric features in the central regions of the galaxies, meaning that the measurements will not be sensitive to low surface brightness features in the galaxy outskirts.  Measuring $r_{70}$ using the 10 pixel total flux aperture does not significantly affect the results as the measurement varies by less than 1 pixel in the majority of galaxies when measured from apertures with 15, 20 or 25 pixel radii.  Although the asymmetry measurements themselves vary a little when determining $r_{70}$ from these different sized total flux apertures, they do not vary by enough to significantly impact the fraction of objects determined to be `disturbed' (where the determination of whether an object is `disturbed' is described in Section \ref{section:asymmetryDistributions}).

\section{Simulations}
\label{section:simulations}

\begin{figure}
  \centering
  \subfigure{\includegraphics[width=3in]{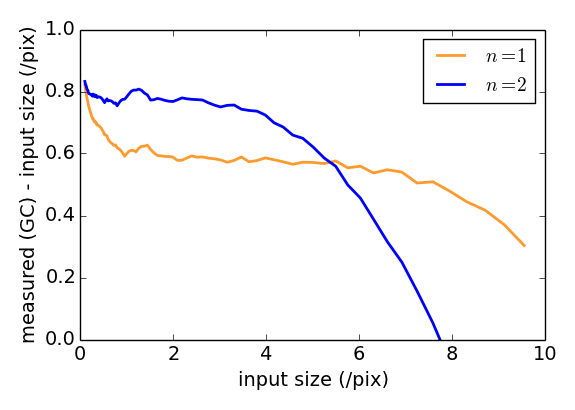}}
  \subfigure{\includegraphics[width=3in]{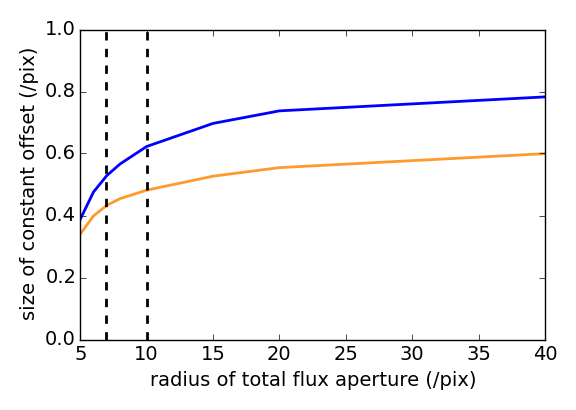}}
  \subfigure{\includegraphics[width=3in]{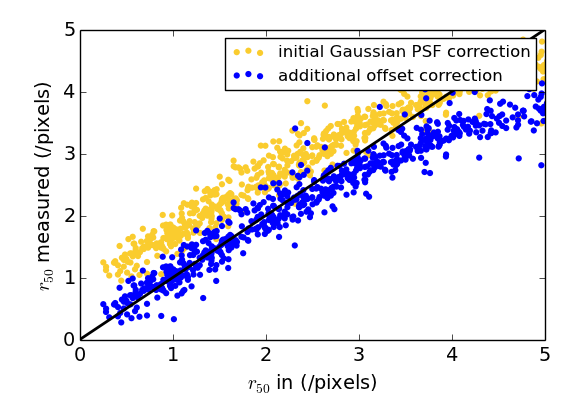}}
  \caption{Showing the nature of the offsets in size produced by the wings in the \textit{HST} PSF for the $i_{775}$ image.  The top panel shows the size difference between the input and measured sizes after an initial PSF correction using a Gaussian approximation (Gaussian correction - GC) for profiles with two different S\'ersic indices, as indicated in the legend (face-on profiles only).  These sizes are measured within a 40 pixel radius total flux aperture.  The offset is approximately constant until the sizes become underestimated due to the fixed total aperture size.  The middle panel demonstrates the dependence of this constant offset on the size of total flux aperture and the S\'ersic index.  The dashed lines demonstrate the size of the 10 pixel radius total flux aperture and the size this corresponds to for an object artificially redshifted from $z\sim4$ to $z\sim8$.  The bottom panel shows the measurements before and after this correction for the bright objects in simulation set I ($n=1$ and with a full range of inclination angles) when using the 10 pixel total flux aperture.}
  \label{fig:psfCalibration}
\end{figure}

\begin{table*}
 \centering
  \caption{Summary of different simulation sets.  Each of the simulations using distributions of single S\'ersic profiles (simulations I, II and III) also allow a uniform range of axis ratios between 0.2 and 1, and a uniform range of total magnitudes between m$_{1500}=23$ and 31.  See text for explanation of log-normal $r_{50}$ parameter choice.}
  \label{simulationsSummary}
  \begin{tabular}{@{}lcccc@{}}
  \hline
  \hline
  Simulation 	& Aim				& \multicolumn{2}{l}{Distribution of parameters:}\\
 ID 			& 				& S\'ersic index ($n$) &$r_{50}$			 \\
 \hline
  I 			& Measurement diagnostics& 1				& 	$0.5 < r_{50} ~\textrm{(/pix)}< 7${ }					\\
  \\
  II			& Typical size biases& 1				&  	log-normal, $\sigma(\textrm{log space}) = 0.2$, $\mu(r_{50}~\textrm{/pix}) = 3.16$	\\
  \\
  III			& Asymmetry measurement distributions & $0.5 < n < 4.5$ & $0.5 < r_{50} ~\textrm{(/pix)}< 10$								\\
 \\
 IV			& Set up null hypothesis from  $z\sim4$ galaxies		&   \multicolumn{2}{c}{N/A - artificially redshifted $z\sim4$ sample}						\\
\hline
\hline
\end{tabular}
\label{table:simulations}
\end{table*}

In this section the different simulations used throughout this paper are described.  There are two different types of simulations, those using simulated single S\'ersic profiles and those employing the artificial redshifting of galaxies in the $z\sim4$ sample (summarised in Table \ref{table:simulations}).

When measuring sizes and morphological disturbance of high redshift galaxies, which are small and faint, it is important to understand the limits of the measurement diagnostics and how they can impact the final results.  The main factors affecting morphology and size measurements are resolution and surface brightness.  Surface brightness depends on both the size and total flux of an object, so simulation set I is designed to investigate how well the CoG algorithm and {\scshape SExtractor} reproduce the sizes of large, faint objects.  Additionally, this simulation set allows for calibration of the total flux measurements and PSF correction.

Simulation set II is concerned with how well the typical sizes of galaxies can be determined in the face of measurement biases. This requires a firm understanding of what we mean by the `typical' size of the population. 

Starting with the assumption that all the selected galaxies are well described by relaxed discs,
the actual disc size is expected to depend on both the virial radius of the parent halo and its spin ($R_{e}\propto\lambda R_{vir}$, where $\lambda$ is the halo spin parameter, $R_{e}$ is the effective radius of the disc galaxy, and $R_{vir}$ is the virial radius of the parent halo).  Halo spins are expected to be distributed log-normally and we can see from our sample (Fig.~\ref{fig:size_logNorm_allRedshifts}) that the galaxy sizes at $z=4-5$ approximate a log-normal distribution. 

As argued by \cite{Huang2013}, if the halo spin parameter is only weakly dependent on redshift and halo mass \citep{Barnes1987,Bullock2001}, then to measure how the typical galaxy size evolves, we need to plot the evolution in the \textit{peak} of this distribution.  Previous studies have plotted the mean galaxy size as a function of redshift (e.g. \citealt{Ferguson2004,Bouwens2004,Oesch2010a,Ono2013}) which can be biased to large sizes due to the tail in the distribution.  Simulation set II is therefore set-up to investigate how well the peak in a lognormal distribution is recovered with different diagnostics, and is used to define firm flux limits above which the peak is accurately reproduced and unaffected by measurement biases.

The final two simulation sets are designed to address the issue of different surface-brightness limits in different images.  Since we always use the image closest in wavelength to the rest-frame $\lambda_{rest}=1500$\AA, objects selected at different redshifts are subject to different image depths.

Simulation set III uses the asymmetry values of simulated single S\'ersic profiles to determine the cut in asymmetry above which an object can be distinguished from a smooth, axisymmetric profile.

Finally, simulation set IV artificially redshifts the $z\sim4$ sample to be used as a test case for null evolution in both sizes and incidence of morphological disturbance of galaxies, so that resolution and surface-brightness effects can be estimated independently of any underlying evolution that we wish to measure within sample.

\subsection{Blank background images}
\label{section:blankBackground}

Each of the simulations (except the artificially redshifted $z\sim4$ sample) employ  blank background images for all relevant filters and surveys into which the simulated galaxies are inserted.  These images were made to mimic the true image background using the following prescription.  First, objects were masked from the real imaging data using a segmentation map produced by {\scshape SExtractor}.  These masked areas were then filled with blank background taken from the actual image by iteratively shifting and rotating the masked image.  When replacing a previously masked area with a new section of background, the noise was scaled according to the local depth measurements of the image.  New depth measurements were made from the blank images to check that no significant additional structure was added to the background from the method used and that the depths matched those of the original image to within 5\%.  Using blank background images ensured that the measured properties were not affected by nearby sources.

\subsection{Simulation set I: Measurement calibration}
\label{section:measurementCalibration}

A set of simulated single S\'ersic profiles ($n=1$) were produced with uniform distributions of parameters described in Table \ref{table:simulations} as well as a uniform range of axis ratios between 0.2 and 1.  These profiles are inserted into the blank background images described above.  The half-light radii were measured with the CoG algorithm as well as with {\scshape SExtractor} \citep{Bertin1996}.  

\begin{figure}
  \centering
  \subfigure{\includegraphics[width=3in]{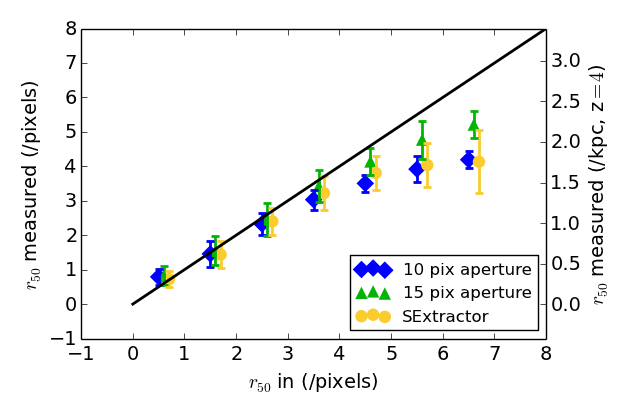}}
  \caption{Results from measuring the half-light radii of single S\'ersic profiles ($n=1$) inserted into the GOODS-S deep $i_{775}$ image with a uniform distribution of sizes and luminosities.  This figure shows the measured half-light radius versus true half-light radius  for all profiles with input total magnitude $m_{tot} < 27$.  The measured sizes are binned according to input size with $\Delta r_{50} = 1$, and the medians and standard deviations are plotted as the points and error bars respectively.  The different measurements plotted are two CoG measurements with different sized total flux apertures; a 10 pixel radius aperture (10 pix aperture, blue diamonds) and a 15 pixel radius aperture (15 pix aperture, green triangles); as well as the half-light radii measured using {\scshape SExtractor} (yellow circles).  Each measurement type has been given a small $x$-axis offset for clarity. The right hand $y$-axis shows the physical sizes that these half-light radii would correspond to at $z\sim4$. }
  \label{fig:diagnosticPlot}
\end{figure}

\begin{figure}
  \centering
  \subfigure{\includegraphics[width=3in]{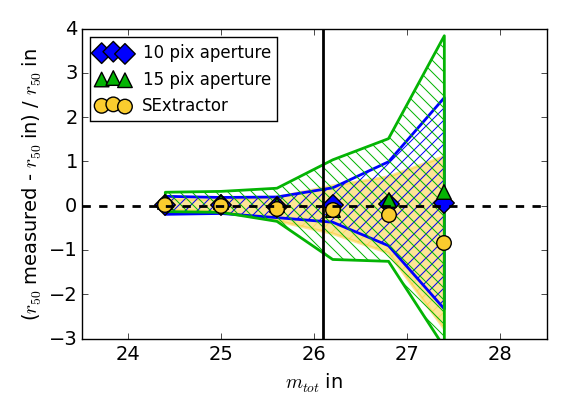}}
  \subfigure{\includegraphics[width=3in]{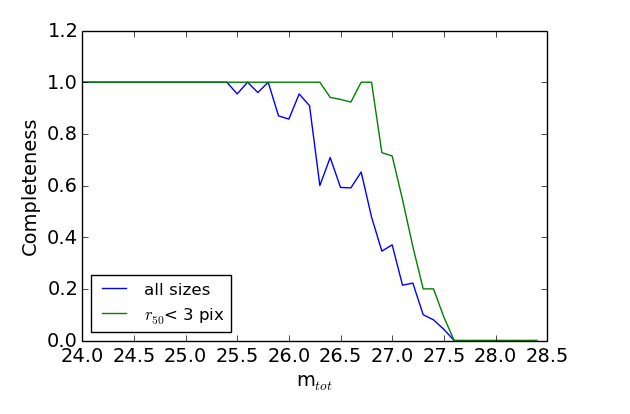}}
  \caption{ \textbf{The top panel} shows the fractional difference between measured and input half-light radii as a function of total magnitude for all simulated galaxies passing the initial $5\sigma$ selection criteria within a 0.3\asec diameter aperture (within the GOODS-S deep $i_{775}$ image).  These results are plotted for all profiles with input $r_{50} <3~\textrm{pixels}$.  The mean values in bins of input magnitude are plotted as points and the shaded regions encompass mean $\pm\sigma$ where $\sigma$ is the standard deviation of the fractional difference in size estimates in the bin. The vertical black line shows the final magnitude cut employed for this image (see Section \ref{section:sizeBias}). The colours and data point symbols correspond to the same measurements as in the first panel. \textbf{The bottom panel} shows the sample completeness (fraction of objects passing the initial $5\sigma$ selection criteria) as a function of total magnitude for all objects in the sample.  The blue line shows the completeness for all selected galaxies and the green for all those galaxies with $r_{50} < 3$, demonstrating that the sample completeness is also size-dependent.}
  \label{fig:diagnosticPlot_size}
\end{figure}

\subsubsection{Measurement calibration}

\textit{PSF correction}
\label{section:psf}
\vspace{0.2cm}

\noindent The initial PSF correction is applied using $r_{50,\textnormal{\tiny{corr}}}=\sqrt{r_{50,\textnormal{\tiny{obs}}}^2-2\textrm{ln}(2)*\sigma_{\textnormal{\tiny{PSF}}}^2}$  where $\sigma_{\textnormal{\tiny{PSF}}}$ is determined from fitting a Gaussian profile to the central region of the PSF.  Sizes are often corrected for PSF effects in quadrature (e.g. \citealt{Oesch2010a}) and the top panel of Fig. \ref{fig:psfCalibration} shows that this is a reasonable approximation for $n=2$ S\'ersic profiles of all sizes as after the initial correction is applied there is an approximately constant offset between input and measured sizes (the deviation at large sizes is due to systematic under-estimation of sizes when using a fixed total flux aperture, see Section \ref{section:measurementBiases}).  The PSF correction performs slightly less well for $n=1$ profiles with  $r_{50}<1$ pixel, although the effect is small, with objects of $r_{50}\sim0.5$ pixels having their sizes over-estimated by $\sim0.1$ pixels (after correction for the constant offset) and so this does not significantly affect the main results. 

The initial PSF correction does not take account of the extended wings in the PSF.  When measuring the total fluxes with large apertures the wings have a substantial impact on the measured sizes.  The wings are much larger in extent than the objects of interest and act to distribute a fixed fraction of the object's flux to large radii. When using fixed sized apertures (larger than the extent of the object itself), the wings generate a constant offset between the measured half-light radius and that of the original profile (after initial quadrature correction for the PSF).  

The measured offsets are of order $\sim0.6-0.9$ pixels for the 10 pixel radius total flux aperture.  These offsets are dependent on the filter used, the size of total flux aperture (what fraction of the total flux is contained within the wings) and the intrinsic profile shape.  These dependencies are displayed in Fig. \ref{fig:psfCalibration}.  Any dependence on intrinsic profile shape is fairly weak compared to the magnitude of the offset.  Changing the simulation profiles from $n=1$ to $=2$ produced a $\sim20-30\%$ change in derived offset.

\begin{figure}
  \centering
  \subfigure{\includegraphics[width=3in]{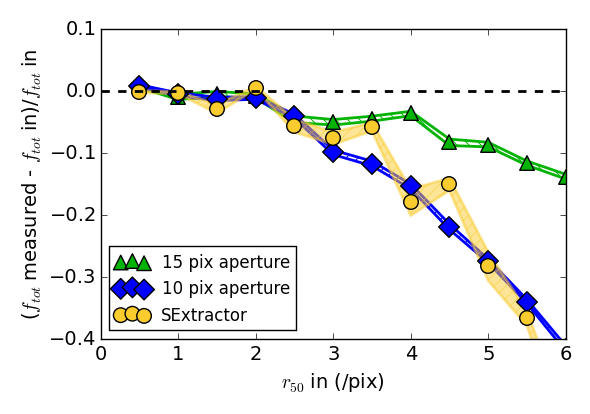}}
  \caption{This plot shows
  the fractional biases in total flux measurements as a function of profile size for the three different measurement techniques.  Only results for profiles inserted into the GOODS-S deep $i_{775}$ image with total magnitude brighter than 25.5 are plotted.  The symbols and shading are plotted as in Fig. \ref{fig:diagnosticPlot}.  The 10 pixel radius aperture underestimates the total flux at smaller sizes than for both {\scshape SExtractor} and the 15 pixel radius aperture.   Total flux estimates using the 15 pixel radius aperture are less biased than {\scshape SExtractor} and the 10 pixel aperture estimates for large profiles.}
  \label{fig:diagnosticPlot_totMag}
\end{figure}

We therefore correct the measured sizes for PSF effects on an image-by-image basis using the offsets derived from single S\'ersic profiles with $n=1$, a full range of inclination angles and the size of total flux aperture used for the final measurements (as shown in the bottom panel of Fig. \ref{fig:psfCalibration}).  These corrections are constant with respect to  intrinsic profile size and are applied after the initial PSF correction has been applied in quadrature.

\vspace{0.5cm}
\noindent\textit{Total flux corrections}
\vspace{0.2cm}

\noindent Total flux corrections are also  required due to the finite sized total flux aperture and are derived directly from input to output total flux measurements of profiles with small input sizes (chosen as objects with measurements that agree well with the input half-light radii).  These corrections are of order 6-10\% within the 10 pixel radius aperture and are applied on an image-by-image basis. Although these corrections have been derived from simulated profiles, they are consistent with the point source total flux corrections derived from stars within the images.

\subsubsection{Measurement biases}
\label{section:measurementBiases}

 The comparison of input to corrected output sizes measured with {\scshape SExtractor} and the CoG algorithm with two different sized total flux apertures are displayed in Fig. \ref{fig:diagnosticPlot}.  The size measurements derived with these different tools  are all underestimated for the largest profiles.  Both {\scshape SExtractor} and the CoG-based measurements using a total flux aperture with a radius of 10 pixels systematically underestimate the sizes of S\'ersic profiles ($n=1$) with physical sizes $r_{50}\sim1.3~ \textrm{kpc}~ (z=4)$ (Fig \ref{fig:psfCalibration} shows that the sizes start to be underestimated at $\gtrsim2.5$ pixels (1 kpc, $z=4$), but not significantly so below sizes of $\sim3$ pixels ($\sim1.3$ kpc, $z=4$)). This scale is not much larger than the mean galaxy size of $\sim1.2$ kpc measured previously at $z\sim4$ \citep{Oesch2010a}.  The results for a slightly larger total flux aperture (with radius of 15 pixels) better reproduce the sizes of the largest galaxies but with a trade-off.  Increasing the aperture size reduces the image depth which in turn increases the noise in the total flux measurements and subsequently in the half-light radii.

 The top panel in Fig. \ref{fig:diagnosticPlot_size} shows the \textbf{fractional} difference between the measured and input half-light radii as a function of input total magnitude for all simulated objects passing the $5\sigma$ detection threshold for this image ($m(\textrm{AB})<28$ within a 0.3\asec diameter aperture). This plot shows that the CoG algorithm with 10 pixel aperture reproduces the sizes of the galaxies well, whereas {\scshape SExtractor} starts to underestimate the sizes of faint galaxies.  The measurements with the 15 pixel aperture are much noisier than those from both the 10 pixel aperture and {\scshape SExtractor}.

The completeness of the sample is determined by the flux within the very central regions (within the original selection aperture) and so will not be uniform across bins in total input magnitude.  This is shown in the bottom panel of Fig. \ref{fig:diagnosticPlot_size}.  The completeness for all objects with $r_{50}<3$ pixels are displayed as well as that of the whole sample, showing that the completeness drops at brighter magnitudes for the largest objects. 
 
A comparison of the total flux measurements taken with the different estimators is displayed in Fig. \ref{fig:diagnosticPlot_totMag} as a function of simulated profile size.  Only objects with total magnitude brighter than 26.1 are used to determine the median and scatter in the offsets.  This plot shows that all the total flux measurements underestimate total fluxes for objects with radii larger than 2 pixels. The 15 pixel aperture provides the best measurement for the total flux with less than 15\% average underestimation of the total magnitude for all simulated profiles and less than 5\% underestimation for profiles with radii $\leq4$ pixels.

\vspace{0.5cm}
These simulations display four
important points:
\begin{itemize}
\item The CoG algorithms performs as well as {\scshape SExtractor} at measuring the sizes of single S\'ersic profiles.
\item Both measurement techniques systematically underestimate the sizes of the largest galaxies, and this needs to be addressed when measuring the typical sizes of high-redshift galaxies from a given size distribution.
\item Both measurement techniques systematically underestimate the total flux of the largest galaxies, but the 15 pixel aperture provides the least-biased total flux estimates.
\item There needs to be a flux limit above that of the original selection limit that minimises the effects of size-dependent completeness and the biases introduced at faint total fluxes.  
\end{itemize}

\begin{figure}
  \centering
  \subfigure{\includegraphics[width=3.1in]{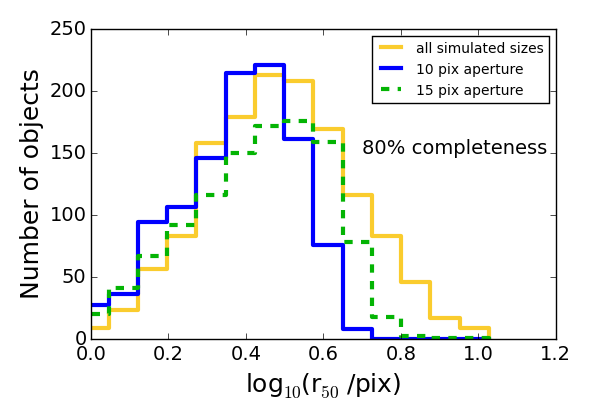}}
  \subfigure{\includegraphics[width=3.1in]{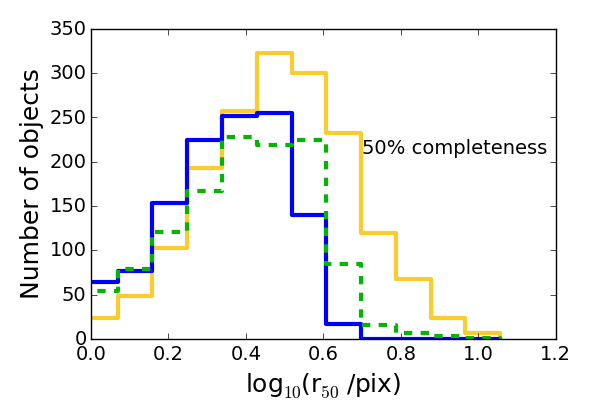}}
  \subfigure{\includegraphics[width=3.1in]{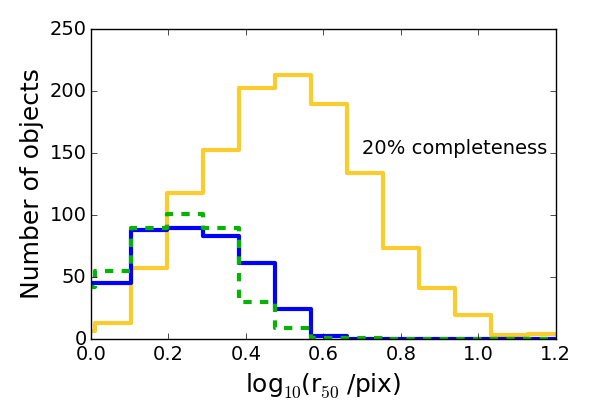}}
  \caption{These histograms display the underlying size distribution in the simulations and the recovered distribution (objects meeting the $5\sigma$ detection threshold within the selection aperture, see text for details) at different completeness cuts.  The simulated objects were inserted into the CANDELS GOODS-S deep $i_{775}$-band image and are plotted according to their input magnitudes and sizes.  The recovered objects are plotted according to the total fluxes measured within the 15 pixel aperture and sizes measured using the 10 pixel aperture.  The objects are taken from bins of width 0.2 mag centered on the magnitude at which the completeness reaches the value quoted.  The greater number of objects within the 50\% cut bin when selected from the 15 pixel aperture are due to the underestimation of sizes of the larger apertures.}
  \label{fig:recoveredSizeDistributions}
\end{figure}

\subsection{Simulation set II: Typical galaxy size bias}
\label{section:sizeBias}

There are two main factors potentially impeding the accurate measurement of typical galaxy sizes given an underlying size distribution: the completeness of the sample and the biases affecting size measurements of the largest profiles. To address these issues we use a simulation consisting of profiles with a lognormal size distribution with mean size and scatter typical of that measured by previous studies at $z\sim4$.  This simulation is designed to determine what sized aperture is sufficient for determining half-light radii given an expected size distribution as well as determining flux cuts, above which the typical size of galaxies can be reliably determined.

  Fig. \ref{fig:recoveredSizeDistributions} shows how the recovered size distributions are affected at different completeness levels.  It is noteworthy that the large-size tail of the galaxy size distribution is not fully recovered, even for very high levels of completeness.  However, our adoption of the modal estimator allows us to recover the typical size for completeness levels $\geq50$\%.

\begin{figure*}
  \centering
  \subfigure[Mean size estimates]{\includegraphics[width=3.4in]{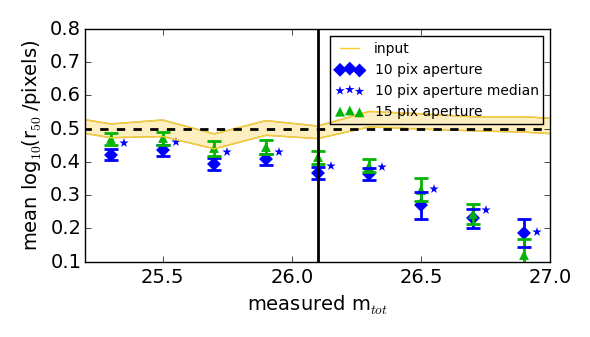}}
  \subfigure[Modal size estimates]{\includegraphics[width=3.4in]{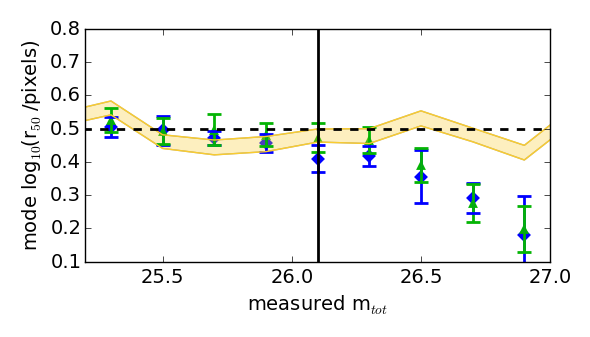}}
  \caption{Typical galaxy size measurements for a simulated set of galaxies inserted into the GOODS-S deep $i_{775}$ blank background image, uniformly distributed in total magnitude.  The simulated size distribution is lognormal with a peak at $\sim3.16$ pixels ($\sim1.3$ kpc, $z=4$) and a width of $\sigma(\textrm{log}_{10}(r_{50} /\textrm{pixels})) = 0.2$. Panel (a) shows the results using the mean to estimate the typical galaxy sizes, while panel (b) shows the results using the mode.  The median results for the 10 pix aperture are also shown as stars in panel (a) for completeness, with small offsets added in the $x$-axis for clarity.   The green and blue points with error bars  show the mean/mode and standard error of the measured sizes, calculated within magnitude bins of width $\Delta 
 m_{tot} = 0.2$, using size measurements derived  from total flux apertures of radius 10 pixels and 15 pixels, respectively. The yellow shaded region shows the input typical sizes and associated errors. The vertical line in each plot shows the magnitude limit chosen for this image (see text for details).  Objects fainter than this magnitude limit are not used for further analysis and the magnitude limit is chosen on an image-by-image basis. This figure demonstrates the validity of adopting the modal size estimator to determine typical galaxy sizes at high redshift.}
  \label{fig:typicalSizeBias}
\end{figure*}

Fig. \ref{fig:typicalSizeBias} shows the typical size measurements made for two different input size distributions, using two different measurement diagnostics; the mode\footnote{The mode is estimated as in {\scshape SExtractor}, using $\textrm{mode}=2.5*\textrm{median}-1.5*\textrm{mean}$ after sigma clipping until convergence around $\textrm{median}\pm3\sigma$} and the mean\footnote{In this case the mean is taken in log space and is not directly comparable to the mean sizes reported in previous studies.}.  In the absence of biases, taking the mean in log-space of the sizes should trace the peak in the size-distribution as required to compare to the Mo et al. (1998) framework and would be preferable to the noisier mode estimator.  These plots show that, for a size distribution expected for galaxies at $z\sim4$, the mean will systematically underestimate the typical galaxy size with both sized apertures, a bias which increases dramatically as the completeness of the recovered sample drops (left-hand panel).  The modal estimator, although noisier, provides less biased typical size estimates using both sized apertures.  For completeness, we also show the results for the median which is marginally less biased than the mean.

Based on the results of these simulations, we therefore choose a magnitude limit, measured within a 15 pixel radius aperture, at which a population with size distribution expected at $z\sim4$ is recovered at 50\% completeness (Table \ref{table:fluxLimits}) and we measure the typical size at each redshift using a modal estimator.

\begin{figure}
  \centering
  \subfigure{\includegraphics[width=3in]{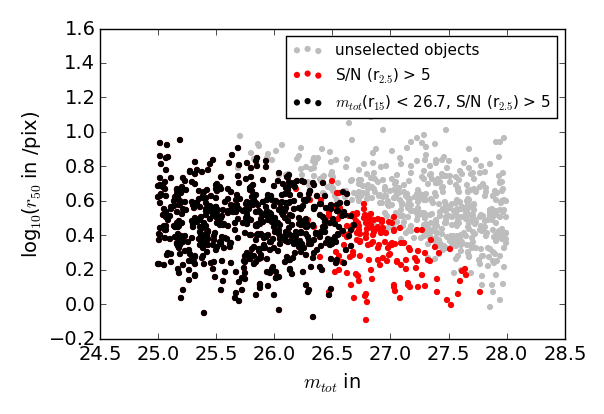}}
  \caption{This plot shows how a bright flux cut imposed based on photometry performed within a large aperture mitigates selection biases. It shows the input sizes and magnitudes of simulated profiles added to the GOODS-S deep $V_{606}$ image.  They are colour-coded depending on what selection criteria they would meet (grey points do not meet any selection criteria).  Those objects satisfying S/N$>5$ within a 5 pixel (0.3\asec) diameter aperture (required for a galaxy to be initially selected into our sample for photometric redshift analysis) are shown as red points and those \textbf{that also meet} the chosen flux cut in the 15 pixel radius total flux aperture are shown as black points (the selection mechanism employed in this paper).  An ideal selection mechanism would impose a vertical cut in this graph, without any under- or over-sampling of large or small galaxies.  Our selection balances this requirement with choosing a flux cut that is faint enough to provide statistically relevant sample sizes.  We therefore require that the flux cut does not bias the selection of galaxies smaller than or equal to the typical size expected at $z\sim4$.  This plot also shows that a selection mechanism employing fluxes measured within a very small aperture will significantly bias the selection of galaxies.}
  \label{fig:selectionBiases}
\end{figure}

Figure \ref{fig:selectionBiases} displays the effect of this selection criteria on the sampling of the underlying distribution within the simulations.  For each galaxy simulated it is coloured, as indicated in the legend, depending on whether it enters into the inital sample and whether it passes the bright flux cut imposed to ensure we can recover the typical sizes.  An ideal selection criterion would produce a vertical cut in this figure, i.e. it would be based on the total flux within the galaxy.  To achieve this given the biases introduced by identifying the galaxies within small apertures (a necessity when selecting high-redshift galaxy samples) we would have to impose flux cuts $\sim1$ mag brighter than those used here.  To provide statistically useful samples, however, we choose a selection mechanism that does not bias the selection of galaxies with sizes smaller than or equal to the typical sizes we expect at $z\sim4$.

These flux limits will prevent biased estimates if the typical sizes evolve as claimed by previous studies seeing that, for size evolution $\propto(1+z)^{-1}$, the typical galaxy size subtended on the sky (in arcseconds) will decrease with redshift.  Within a scenario where the typical sizes do not evolve (the scenario that we would hope to be able to reject with the current data),  typical sizes of $\sim1.3$ kpc (as measured at $z\sim4$) would correspond to sizes of $\sim4.5$ pixels at $z\sim8$, which would mean that even galaxies of the typical size would be underestimated at the highest redshifts.  These apertures and flux limits are therefore insufficient to prevent unbiased estimation of sizes of galaxies in the highest redshift bins in the absence of size evolution and we take this into account in the null hypothesis test described in Section \ref{section:nullHypothesis}.

\begin{table}
  \centering
  \caption{The magnitude limits imposed on an image-by-image basis.  Objects with total magnitudes fainter than these limits are excluded from further analysis.  These limits have been determined from measurements recovered from simulated single S\'ersic profiles with a lognormal size distribution (see Section \ref{section:sizeBias} for more details).}
  \begin{tabular}{@{}rcc@{}}
  \hline
  \hline
  Field & Filter & Size measurement\\
        &        & magnitude limit\\
  \hline
  GOODS-S      & $V_{606}$ & 26.7 \\
  deep		   & $i_{775}$ & 26.1 \\
               & $z_{850}$ & 26.3 \\
               & $Y_{105}$ & 26.7 \\
               & $J_{125}$ & 26.5 \\
               & $H_{160}$ & 26.3 \\
  \hline
  GOODS-S      & $V_{606}$ & 26.7 \\
  ERS		   & $i_{775}$ & 26.1 \\
               & $z_{850}$ & 26.3 \\
               & $Y_{098}$ & 26.1 \\
               & $J_{125}$ & 26.5 \\
               & $H_{160}$ & 26.1 \\
  \hline
  GOODS-S      & $V_{606}$ & 26.7 \\
  wide		   & $i_{775}$ & 26.1 \\
               & $z_{850}$ & 26.3 \\
               & $Y_{105}$ & 25.7 \\
               & $J_{125}$ & 25.9 \\
               & $H_{160}$ & 25.5 \\
  \hline
          HUDF & $V_{606}$ & 28.5 \\
  			   & $i_{775}$ & 28.1 \\
               & $z_{850}$ & 27.7 \\
               & $Y_{105}$ & 28.5 \\
               & $J_{125}$ & 28.1 \\
               & $J_{140}$ & 28.1 \\
               & $H_{160}$ & 28.1 \\
  \hline
       HUDF-p1 & $i_{775}$ & 27.1 \\
               & $z_{850}$ & 27.1 \\
               & $Y_{105}$ & 27.3 \\
               & $J_{125}$ & 27.5 \\
               & $H_{160}$ & 26.9 \\
  \hline
       HUDF-p2 & $i_{775}$ & 27.1 \\
               & $z_{850}$ & 27.3 \\
               & $Y_{105}$ & 27.3 \\
               & $J_{125}$ & 27.5 \\
               & $H_{160}$ & 27.1 \\
  \hline
  \hline
\end{tabular}
\label{table:fluxLimits}
\end{table}

\subsection{Impact of S\'ersic index $n=1$ choice in simulations I and II}

The choice of a single S\'ersic index of $n=1$ in the measurement diagnostic and typical size bias simulations was investigated by repeating the simulations with $n=2$ (All plots are presented in the Appendix).  All of the measurements (CoG with 10- and 15 pixel apertures and {\scshape SExtractor}) start to systematically underestimate the sizes of the largest profiles at smaller sizes than for profiles with $n=1$, with the 10 pixel apertures and {\scshape SExtractor} underestimating sizes for an input half-light radius of 2 pixels ($\sim0.9$ kpc at $z\sim4$).  This leads to the modal size estimates from a 10 pixel aperture in the size bias simulations (II) being biased to low values for the larger input sizes, the 15 pixel modal estimates remain unbiased.  The flux cuts for reliable typical size estimates remain unchanged.  We therefore test the main results of this analysis with the 15 pixel apertures to test whether the modal values of the true samples differ from those derived with the 10 pixel aperture (see Section \ref{section:size_evolution}). 

\subsection{Simulation set III: Asymmetries of smooth profiles}
\label{section:asymmetryDistributions}

\begin{figure}
  \centering
  \includegraphics[width=3in]{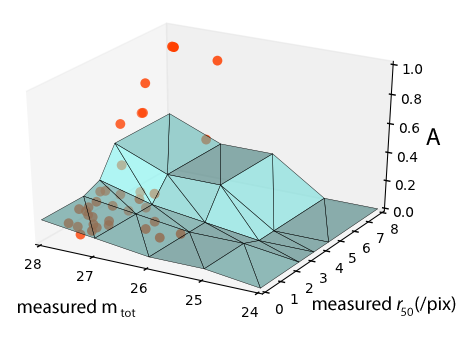}
  \caption{Measured asymmetries (A) as a function of measured total magnitude (m$_{tot}$) and size in the HUDF $i_{775}$ image.  The orange points show the measured asymmetries for galaxies that have their properties measured within this image.  The blue surface shows the asymmetry value below which the measurements of 98\% of the simulated single S\'ersic profiles reside (see text for details).  Objects lying above this surface are labelled as `disturbed' and objects below the surface cannot be distinguished from smooth, axisymmetric profiles.}
  \label{fig:asymmetrySimulations}
\end{figure}

To determine whether the measured asymmetry values for the selected objects are consistent with the objects being smooth and symmetric, they are compared to the measurements derived from a large set of single S\'ersic profiles inserted into realistic background images (see Section \ref{section:blankBackground}).  Asymmetry values for smooth profiles inserted into true images deviate from zero primarily due to pixelation, centering and image noise.  For objects selected at the redshifts studied in this paper these effects are large as the galaxies are small and faint.

Single S\'ersic profiles were added to the blank background images with a wide range of fluxes, half-light radii and S\'ersic indices (see Table~\ref{table:simulations}).  The distribution of measured asymmetries is then used to determine the probability that a galaxy with measured asymmetry, A, is disturbed.  This is simply determined from the fraction of simulated objects, matched in size and flux, that have an asymmetry value smaller than A.  For the following analysis the asymmetry value used to define a disturbed profile is chosen for a probability of $A(1-\textrm{P}(\textrm{Symm} |f_{tot},r_{50})=0.98)$.  In other words, 98\% of the simulated, smooth profiles with matching flux and size have measured asymmetry values lower than the chosen value (see Fig. \ref{fig:asymmetrySimulations}). These simulations allow us to determine whether the asymmetry measurement derived from a real object can be distinguished from that of a smooth profile and to what confidence.  The actual value of 98\% is chosen to be conservative but checks have been made to make sure that any conclusions do not depend on the precise value of $A(1-\textrm{P(Symm)})$ chosen. 

Fig. \ref{fig:asymmetrySimulations} displays how the range of measured asymmetries for smooth profiles depends on their total flux and size (and hence their surface brightness).  The plot shows objects measured within the HUDF $i_{775}$ filter compared to the surface below which 98\% of the simulated single S\'ersic profiles lie.  Objects with asymmetry values higher than the surface are labelled as `disturbed'.  As the surface brightness of the simulated smooth profiles decreases (by decreasing their flux or increasing their size), the noise in the asymmetry values increases and higher asymmetry values are measured (hence the surface rises at large sizes, faint magnitudes). At small sizes the objects become partially unresolved, the asymmetry values are mainly determined by the shape of the image PSF and the measured asymmetries are unlikely to lie above the surface.

\begin{figure*}
  \centering
  \includegraphics[width=6in]{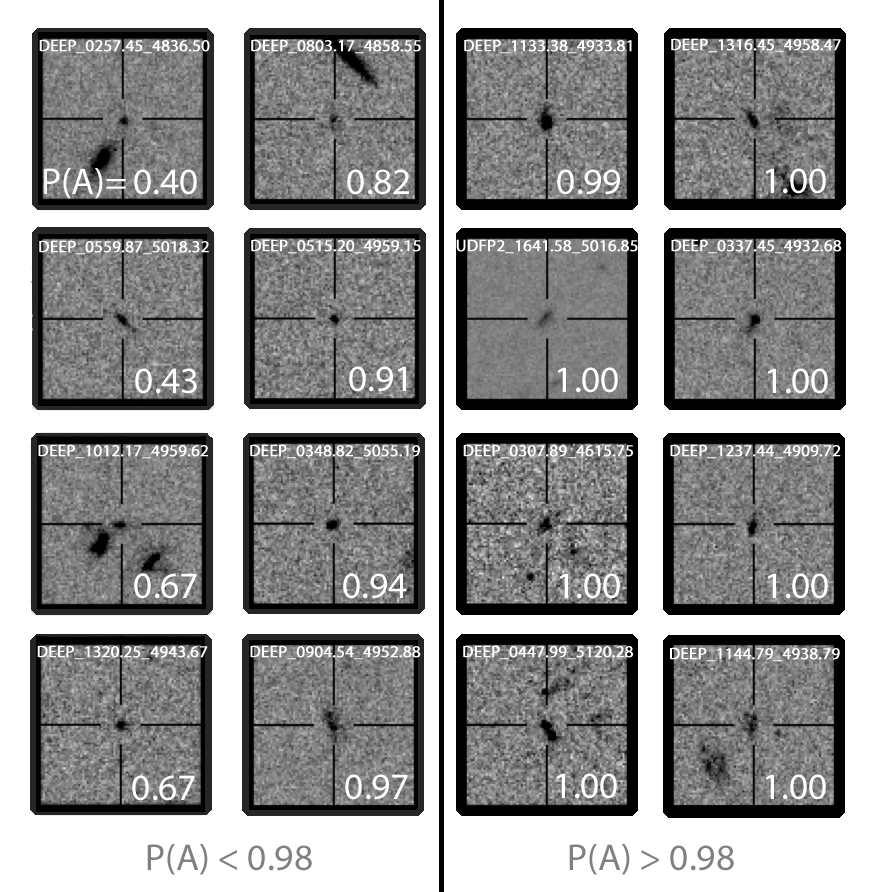}
  \caption{Postage stamps of a selection of \MUV{ }$<-20$ objects in the $z\sim4$ redshift bin separated using the $\textrm{P}(\textrm{A})>0.98$ cut, where we require 98\% of the simulated single S\'ersic profiles have asymmetry values lower than the measured Asymmetry value to be able to label that object as `disturbed'. }
  \label{fig:postageStamps}
\end{figure*}

Postage stamps of a sub-sample of 16 \MUV{ }$<-20$ galaxies are displayed in Fig. \ref{fig:postageStamps}, sorted by the probability that they are disturbed, and separated into objects falling above and below the 0.98 probability cut.  The objects were chosen to demonstrate what types of features contribute to labelling a galaxy as `disturbed', as well as demonstrating what features are present in objects labelled as `smooth', either due to the choice of probability cut (see object DEEP\_0904.54\_4952.88 with P(A) = 0.97) or lack of sensitivity to low-surface-brightness features in the outskirts of the galaxies (see object DEEP\_0803.17\_4858.55).

\subsection{Simulation set IV: Artificially redshifted galaxies}
\label{section:AR}

An artificially redshifted $z\sim4$ galaxy sample is used as a null evolution test case for both sizes and morphologies.  The sample is subjected to the same measurement algorithms and brightness cuts as the true sample in each redshift bin, thereby providing a test for the significance of any measured evolution.

The sample of galaxies at $3.5<z<4.5$ is artificially redshifted into the different redshift bins at $z\sim5$, 6, 7 and 8, in a similar fashion to the method employed in \cite{Bouwens2004a}. For each higher redshift bin, the original sample is randomly assigned a new redshift within the $\Delta z\pm0.5$ interval.  The measurement images chosen for the AR galaxies are those providing wavelength coverage closest to $\lambda_{rest}=1500$\AA~at the new redshift.

The objects are scaled in flux to account for cosmological dimming, and resampled to account for the change in angular diameter distance between the actual redshift and the new assigned redshift.  If the original, re-scaled PSF is expected to be $<90\%$ of the full width at half-maximum (FWHM) of the PSF of the new measurement image then the low-redshift image is Gaussian broadened to match the FWHM of the destination image.  This situation is infrequent as the angular diameter sizes actually increase at these redshifts, and it is only when the destination image FWHM is significantly wider than the original image that any broadening is required.  The re-sampled, scaled object is then inserted into a blank region of the destination measurement image.  It is assumed that the background noise is dominant and so no attempts are made to scale the source Poisson noise counts.

When performing the PSF corrections on AR objects, the initial Gaussian approximation is performed based on the width of the destination image PSF but the correction for the wings in the PSF must be applied using the calibration derived for the original measurement image (as the PSF broadening does not take into account the differing structure in the wings of the two PSFs).

These AR galaxies then have their half-light radii and asymmetries measured using the same methods as used on the actual sample.  Any apparent evolution in any of the derived parameters can then be tested against this sample to ensure that it is not introduced by differences in resolution, sensitivity or selection limits.

\subsection{Summary}

The simulations show the following:

\begin{itemize}
  \item Large galaxy sizes are systematically underestimated by the CoG algorithm used here and by {\scshape SExtractor}. However, the results from the CoG algorithm are more robust at faint magnitudes, where {\scshape SExtractor} estimates can become systematically underestimated.
  \item Although a larger aperture for the CoG algorithm is slightly less biased at large sizes, the measurements are much noisier.
\item The total fluxes are systematically underestimated by all estimators studied but the 15 pixel radius aperture shows the best performance, with the total fluxes not being underestimated by $<5$\% at the expected peak of the $z\sim4$ size distribution.
  \item The typical sizes of galaxies are well reproduced at $z\sim4$ using a modal estimate of the peak in the distribution and the 10 pixel total flux aperture. The mean, however, is biased to small sizes when using the 10 pixel total flux aperture.
  \item To recover the typical size of the underlying population, high sample completeness is required. Strict flux limits are therefore imposed at which the recovered completeness of the expected $z\sim4$ size-distribution is greater than 50\%.  These limits are derived on an image-by-image basis using simulation set II.
  \item If the typical sizes of galaxies are as small as previously measured the CoG algorithm can reproduce these sizes, and does not bias the measurement of the \textit{typical} galaxy size with the flux limits imposed.
\end{itemize}

Based on the simulations performed we base the following work on size measurements using the 10 pixel radius aperture.  This aperture provides less noisy size estimates than the 15 pixel aperture and will still recover the sizes of $z\sim4$ galaxies.  The total fluxes that we report, however, are measured within the 15 pixel aperture to prevent biasing the flux estimates for galaxies at the peak of the size-distribution.  The 15 pixel aperture provides total fluxes that are less biased than both the 10 pixel aperture and {\scshape SExtractor}.  Taking the size and total flux measurements from different sized apertures in this way also ensures that these two measurements are decoupled and the errors are independent (as different annuli are used for determining the background subtraction).

When measuring the typical sizes of galaxies we use a modal estimate, rather than the mean of the distribution to avoid being biased due to the systematic underestimation of the sizes of the largest galaxies as well as the lower completeness of the largest galaxies.  We do not use any objects with total fluxes fainter than the limits given in Table \ref{table:fluxLimits} to avoid biasing the measurement of the peak of the derived size distribution.

\section{Results}

\subsection{Size-Luminosity relation}
\label{section:size-lum}

\begin{figure*}
  \centering
   \subfigure{\includegraphics[width=3in]{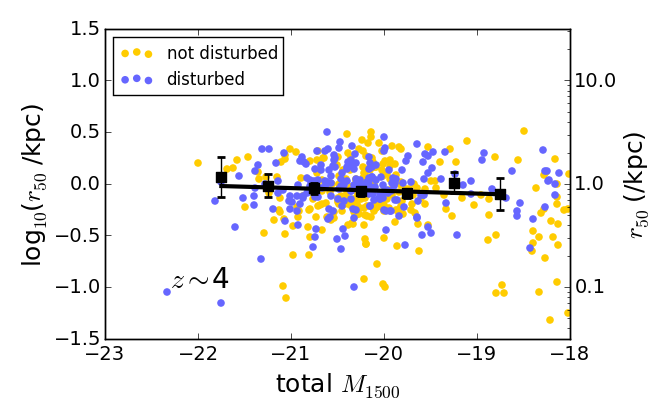}} 
   \subfigure{\includegraphics[width=3in]{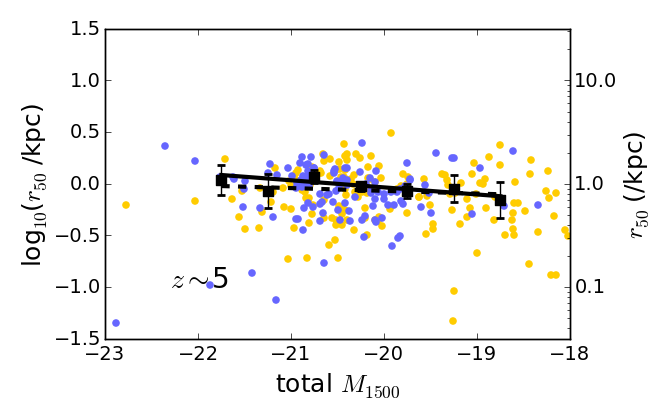}} 
   \subfigure{\includegraphics[width=3in]{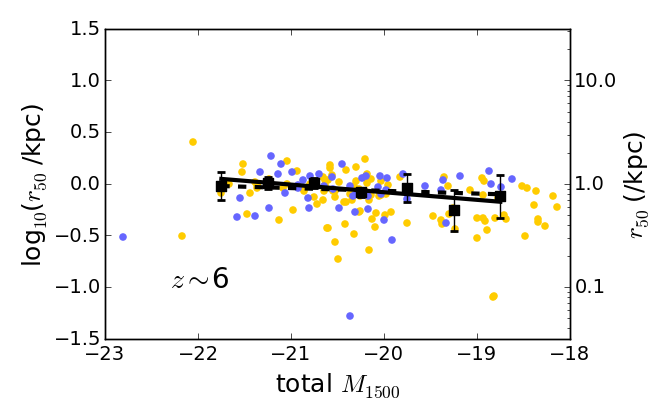}} 
   \subfigure{\includegraphics[width=3in]{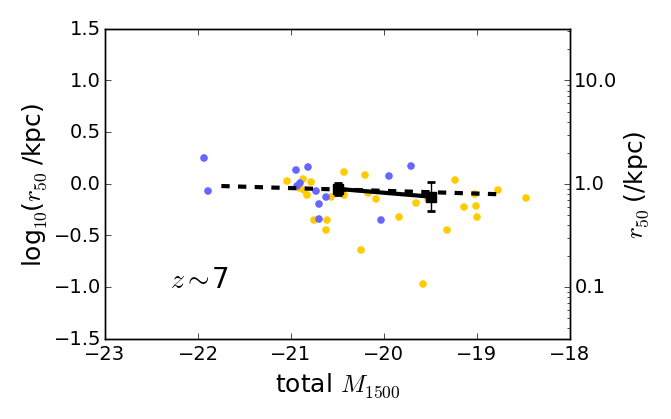}} 
   \subfigure{\includegraphics[width=3in]{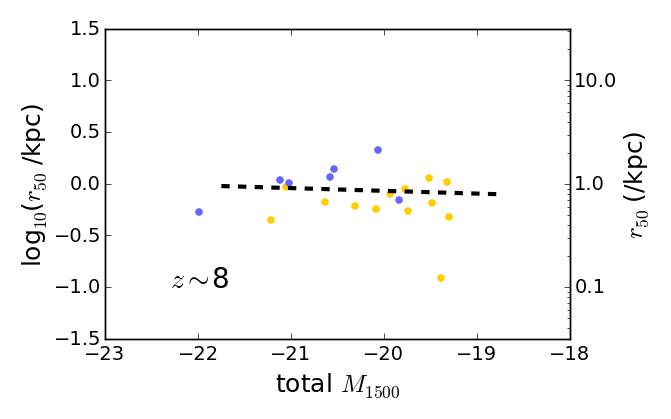}} 
  \caption{Log$_{10}(r_{50}~ \textrm{/kpc})$ versus absolute magnitude in each redshift bin.  The purple points show galaxies with asymmetry values that indicate disturbed profiles and the yellow points have asymmetry values that cannot be distinguished from those derived from axisymmetric single S\'ersic profile fits, matched in UV luminosity and size (see text for details).  The chosen probability cut to distinguish disturbed profiles in this plot is 0.98, i.e. only 2\% of the distribution of simulated axisymmetric profiles have asymmetry values higher than the chosen cut.  The black points show the modal sizes in bins of luminosity with $\Delta{M}_{1500}=0.5$ in the $z\sim4$, 5 and 6 redshift bins, and $\Delta{M}_{1500}=1$ in the $z\sim7$ redshift bin.  There are insufficient objects at $z\sim8$ to allow for calculation of the mode in two separate bins.  The solid black line shows the best-fitting size-luminosity relation in each redshift bin and the dashed black line shows the $z\sim4$ relation for comparison.}
  \label{fig:size_lum_z_bins}
\end{figure*}

The logarithm of the galaxy size is plotted against absolute magnitude in Fig. \ref{fig:size_lum_z_bins}, separated into separate redshift bins with $z\sim4,5,6,7$ and $8$.  The sizes were measured within the 10 pixel radius aperture and are colour coded according to whether they are measured as disturbed or not (see Section \ref{section:asymmetryDistributions}).  The typical sizes of galaxies are measured in bins of width $\Delta$\MUV$\,=0.5$ for the three lowest redshift bins and bins of width $\Delta$\MUV$\, = 1$ at $z\sim7$.   Bootstrap resampling is used to estimate the modal size and the associated uncertainties.  Linear regression is then used to measure the gradient and intercept of the relation in each bin.  There are insufficient objects at $z\sim8$ to provide robust modal estimates within two magnitude bins so the size-luminosity relation is not measured in the highest redshift bin.

\begin{figure}
  \centering
  \subfigure{\includegraphics[width=3in]{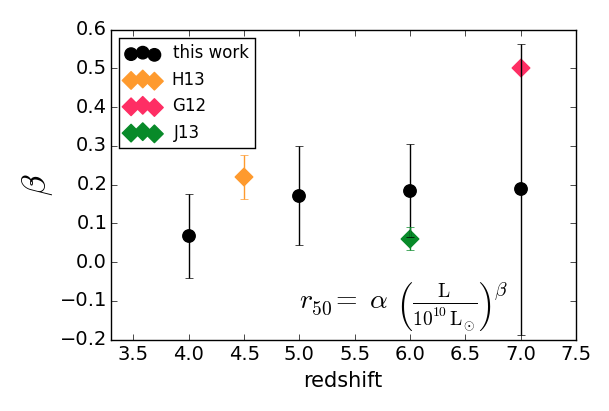}}
  \subfigure{\includegraphics[width=3in]{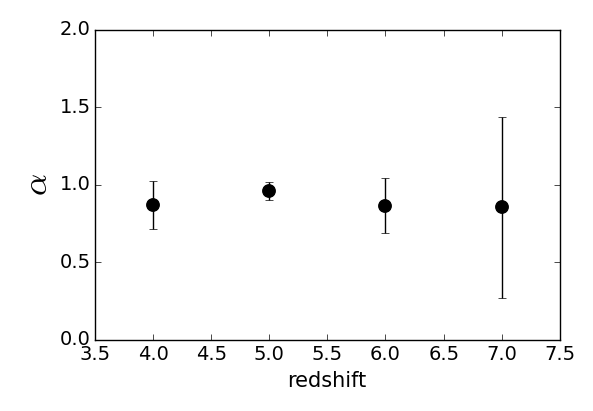}}
  \caption{Size-Luminosity relation plotted as a function of redshift.  The top panel shows the evolution in the exponent of the size-luminosity relation whereas the bottom panel shows the evolution in the normalisation (see text for details).  Previous measurements of the exponent are also plotted in the top panel as indicated in the legend.  H13 refers to the measurements presented at $z\sim4-5$ in \protect\cite{Huang2013}, J13 refers to the $z\sim6$ size-luminosity relation measured for \MUV{ }$<-20$ galaxies in \protect\cite{Jiang2013a} and G12 refers to the $z\sim7$ size-luminosity relation presented in \protect\cite{GrazianA.2012}.}
  \label{fig:size_lum_evoln}
\end{figure}

The evolution of the size-luminosity relation is plotted in Fig. \ref{fig:size_lum_evoln}.  We plot both the evolution in the exponent and normalisation of the relation, where these values are related to the measured gradient and intercept according to equations $2-5$.

\begin{equation}
\textrm{log}_{10}(r_{50})=a\textrm{M}_{1500}+b
\end{equation}
\begin{equation}
r_{50}=\alpha\left(\frac{\textrm{L}}{10^{10}\textrm{L}_\odot}\right)^\beta
\end{equation}
\begin{equation}
\alpha=10^{b\,-\,20.23a}
\end{equation}
\begin{equation}
\beta=-2.5a
\end{equation}

We see no evidence of any evolution in the size-luminosity relation across this redshift range.  The size-luminosity relation is quite shallow in each redshift bin but the errors in the normalisation and exponent, $\alpha$ and $\beta$, are large due to the noisy modal estimates at the extremes in luminosity for all redshift bins.  The results displayed in Fig. \ref{fig:size_lum_evoln} are perfectly consistent with lack of evolution in the relation.  It is also important to note that we have not been able to sufficiently correct for the underestimation of the sizes of the largest galaxies.  Although the distribution of measured sizes at $z\sim7$ and 8 are not distributed evenly about the $z\sim4$ size distribution overplotted on each panel, the scale at which sizes of galaxies start to become underestimated is at $\sim9$kpc at $z\sim8$ (see Fig. \ref{fig:size_logNorm_allRedshifts}) and so the distribution may be biased at sizes larger than the typical sizes of $z\sim4$ galaxies (i.e. above the $z\sim4$ size-luminosity relation).

Given these uncertainties, we find good agreement with the size-luminosity relation measured by \cite{Huang2013}.  The steep size-luminosity relation reported by \cite{GrazianA.2012} at $z\sim7$, although in agreement within our measurement errors,  is not well reproduced by these data.  Although the number of objects to measure the relation at $z\sim7$ in this work is much smaller than in the \cite{GrazianA.2012} analysis, this is due to the imposed flux cuts chosen  to ensure the accurate reproduction of  galaxy sizes.  Without these cuts, and with {\scshape SExtractor} based sizes, our measured size-luminosity relation would steepen significantly due to the preferential inclusion of smaller objects at a given total magnitude at completeness levels $<50\%$ (see Fig. \ref{fig:recoveredSizeDistributions}).

\begin{table}
  \centering
  \caption{A table giving the measured normalisation ($\alpha$) and exponent ($\beta$) of the observed size-luminosity relation for the redshift bins spanning $4\geq{z}\geq7$.}
  \begin{tabular}{@{}ccc@{}}
  \hline
  \hline
  $z$ & $\alpha$ & $\beta$\\
  \hline
  4 & \textbf{$0.86\pm0.15$ }& \textbf{$0.06\pm0.11$} \\
  5 & \textbf{$0.96\pm0.06$} & \textbf{$0.17\pm0.13$} \\
  6 & \textbf{$0.86\pm0.18$} & \textbf{$0.18\pm0.12$}\\
  7 & \textbf{$0.85\pm0.56$} & \textbf{$0.19\pm0.38$} \\
  \hline
  \hline
\end{tabular}
\label{table:sizeLum}
\end{table}

\subsection{Size evolution}

\subsubsection{Lognormal size distribution}

\begin{figure}
  \centering
  \includegraphics[width=3in]{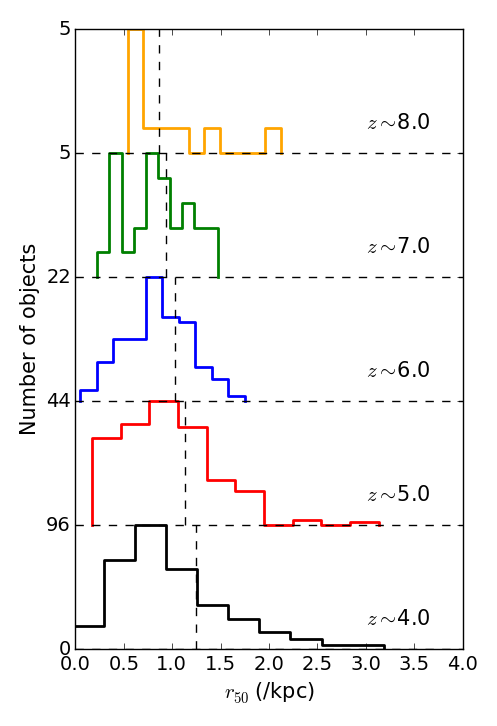}
  \caption{The size distribution of galaxies in the bright luminosity bin ($(0.3-1)\textrm{L}_{*,z=3}$) plotted for each each redshift bin (offset in the y-axis for clarity).  The vertical dashed lines show the approximate scale above which the size estimates are expected to begin to be systematically underestimated ($r_{50}\sim4$ pixels).}
  \label{fig:size_logNorm_allRedshifts}
\end{figure}

The size distribution of galaxies is plotted in separate redshift bins for all bright objects (0.3-1L$_{*(z=3)}$) in Fig. \ref{fig:size_logNorm_allRedshifts}.  We see that the size distribution approximates a lognormal size distribution in the lowest redshift bins.  \cite{Oesch2010} suggest that the size evolution they measure is dominated by the build-up of the tail to large sizes at low redshifts.  Overplotted on the histogram for each redshift bin is the physical scale at which the sizes of galaxies are systematically underestimated with the CoG algorithm, using the 10 pixel total flux aperture (at $r_{50}\sim4$ pixels).  It is not clear from this plot alone to what extent these biases affect the measured size distribution, and this is tested further in Section \ref{section:nullHypothesis}.

As argued at the beginning of Section \ref{section:simulations}, to be able to compare to the theoretical size evolution predictions, however, we need to be tracing the evolution in the \textit{peak} of the size distribution function.  The peak shows little evolution from this plot and this is investigated further using a modal estimator in the following sections.   

\subsubsection{Measured evolution in bright $(0.3-1 \textrm{L}_{*(z=3)})$ and faint $(0.12-0.3 \textrm{L}_{*(z=3)})$ luminosity bins}
\label{section:size_evolution}

\begin{figure}
  \centering
  \includegraphics[width=3.3in]{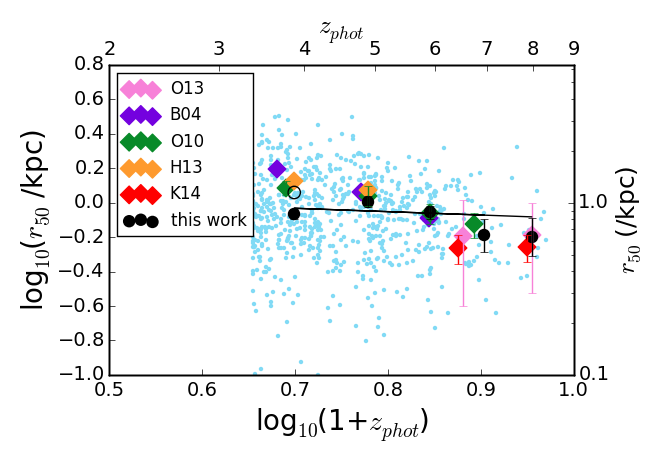}
  \caption{The size evolution at constant luminosity in the range (0.3-1) L$_{*(z=3)}$ ($-21 <\,$\MUV$\,< -19.7$).  The data has been plotted so that the evolution in $r_{50}$ as a function of $(1+z)^n$ can be fit with a straight line.  The values plotted for this work are the mode and the associated errors estimated from bootstrap resampling (see text for details).  Errors are plotted for all redshift bins, but they are smaller than the size of the points in the lowest redshift bins.  The black line shows the best-fit line through the data points with gradient $n=-0.20\pm0.26$, suggesting evolution of $r_{50}\propto(1+z)^{-0.20}$.  Values from the literature are over-plotted with values from \protect\cite{Bouwens2004a} (B04) in purple, those from \protect\cite{Oesch2010a} (010) in green, \protect\cite{Ono2013} (O13) in pink, \protect\cite{Huang2013} (H13) in orange and \protect\cite{Kawamata2014} (K14) in red.  The individual object measurements from this work are plotted in light blue.  The points from B04 and O10 are taken from the mean of the distribution of sizes (rather than the distribution in log space), within the same luminosity bin, whereas the results from H13 are taken from the peak of the distribution (i.e. the mean of the distribution in log space) at \MUV$\,=-21$.  For completeness, we also plot the mean value of measured sizes (in real space) within the 15 pixel aperture for $z\sim4$ (open circle).}
  \label{fig:size_redshift_bright}
\end{figure}

\begin{figure}
  \centering
  \includegraphics[width=3.3in]{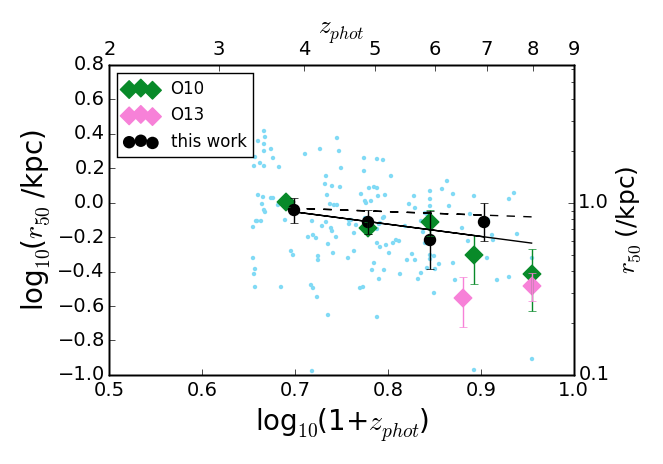}
  \caption{The size evolution at constant luminosity in the range (0.12-0.3) L$_{*(z=3)}$ ($-19.7 < $\MUV$ < -18.6$). Individual size measurements, typical size measurements per luminosity bin and measured evolution are plotted as in Fig. \ref{fig:size_redshift_bright}.  The measured evolution in this luminosity range has a best-fitting gradient of $n=-0.47\pm0.62$. The dashed line shows the fit to the bright luminosity bin.  Values from \protect\cite{Oesch2010a} (O10) are plotted in green and \protect\cite{Ono2013} (O13) are plotted in pink.}
  \label{fig:size_redshift_faint}
\end{figure}

Galaxy size as a function of redshift is plotted in two different luminosity bins, presented in Fig. \ref{fig:size_redshift_bright} (the bright luminosity bin, with $0.3-1 \textrm{L}_{*(z=3)}$) and Fig. \ref{fig:size_redshift_faint} (the faint luminosity bin, with $0.12-0.3 \textrm{L}_{*(z=3)}$).  The logarithm of the typical galaxy size, and associated uncertainties, for each redshift bin is estimated from the modal value with bootstrap resampling of the population.  The best-fitting evolution from these typical sizes is measured using linear regression, incorporating the measurement uncertainties into the fitting.  The gradient from this fit gives the exponent, $n$, in the $r_{50}\propto(1+z)^n$ relationship.  In the bright luminosity bin we measure a gradient of $n=-0.20\pm0.26$ whereas in the faint luminosity bin we measure a gradient of $n=-0.47\pm0.62$.  Both of these measurements are shallower than that expected for a constant halo mass selection, although not significantly so in the faint bin.

The figures show size measurements derived from 10 pixel radius apertures.  The 15 pixel radius apertures that were used to check that the typical size measurements are not significantly biased to smaller values.  The simulations show that, for galaxies with a typical size similar to that measured by \cite{Oesch2010}, both sized apertures  should reproduce the typical size estimate well, and the bins with the best sampling of the underlying population have good agreement between the typical sizes measured with the 10 pixel and 15 pixel aperture.  However, if we measure the mean of the sizes measured at $z\sim4$ within the 15 pixel aperture (not in log-space, in order to replicate the measurements of other studies), we measure a higher value consistent with other studies (shown as an open circle in Fig. \ref{fig:size_redshift_bright}).  If we replace the $z\sim4$ measurement with this value, we would find steeper evolution, with $n=-0.92\pm0.26$, which is in much better agreement with previous measurements presented in the literature.  This comparison could not be made using a mean estimate with the 10 pixel aperture because the systematic underestimation of the sizes of the largest galaxies biases this measurement.  The 15 pixel aperture also gives larger typical size estimates (using the modal estimate) in the two highest-redshift bins ($z\sim7,8$), giving a shallower measured evolution than that measured only with the 10 pixel aperture (gradient of $n=0.08\pm 0.34$), although it is consistent within the errors.

The sampling of the distribution is poor at all redshifts in the faint luminosity bin.  The number of galaxies is too small in the $z\sim8$ bin to allow for bootstrap resampling, so it is not included in the linear regression.   In fact, the typical galaxy sizes are poorly constrained for all redshifts $z>4.5$ in the faint luminosity bin and the associated evolution in sizes is also extremely uncertain.

We test the possibility that the derived size evolution is affected by low-redshift interlopers by measuring the size evolution for firm high-redshift candidates only in the bright bin (see section \ref{section:Selection} for description of firm candidate).  Uncertainties in the derived estimates increase significantly due to poorer sampling of the underlying population, and the derived evolution is shallower, hence low-redshift interlopers are unlikely to be diluting the observed size evolution.

\subsubsection{Comparing to the null hypothesis}
\label{section:nullHypothesis}

There are two effects that make the measurement of the derived evolution uncertain: undersampling of the underlying population by the data and underestimation of the sizes of the largest galaxies.  These effects are not sufficiently taken into account in the uncertainties in the measured gradient.   Using the artificially redshifted $z\sim4$ sample, we can replicate the sampling of the distribution in the highest redshift bins under the assumption of no evolution in galaxy sizes.  This allows us to test the significance of our measured size evolution in the bright luminosity bin without relying on the error in the gradient alone.

The $z\sim4$ sample is artificially redshifted into each redshift bin as described in Section \ref{section:AR}.  Objects are then randomly selected from the AR sample, matching the number of galaxies within each redshift bin.  The gradient of the AR sample (including the original $z\sim4$ population) is then measured employing the same method as applied to the original sample. This is repeated many times to characterise the uncertainties in the derived evolution given the sampling of the distribution provided by the data.

\begin{figure}
  \centering
  \subfigure[Size evolution from $z\sim4$ sample in the (0.3-1)L$_{*(z=3)}$ luminosity bin]{\includegraphics[width=3in]{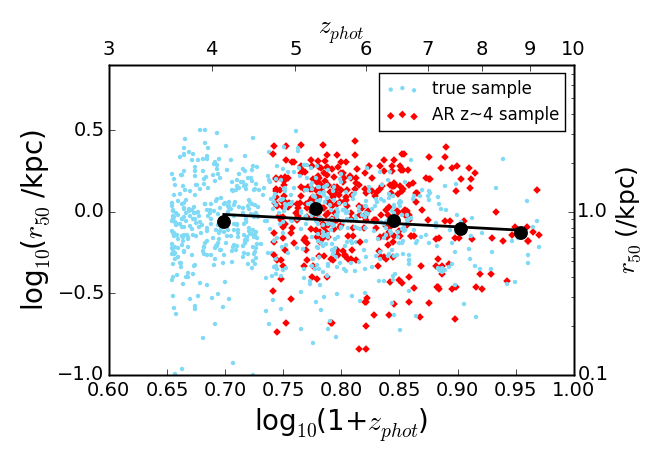}}
  \subfigure[Distribution of measured size evolution from null hypothesis compared to actual measured size evolution.]{\includegraphics[width=3in]{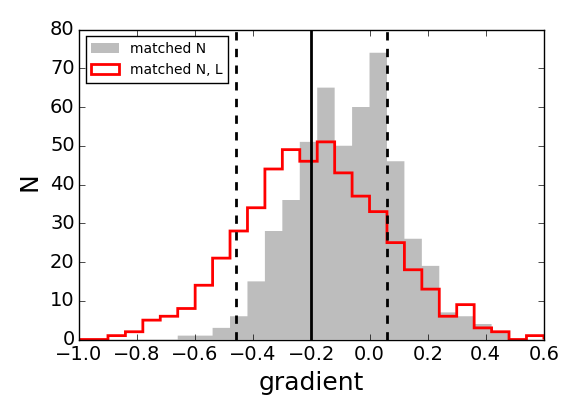}}
  \caption{The results from testing against the null hypothesis.  The $z\sim4$ sample is artificially redshifted into each of the redshift bins, sub-sampled to match the sampling of the distribution by the actual sample (in number), and the gradient measured (see text for more details).  \textbf{The top panel} (a) shows an example of the sampling of the measurements from the AR sample (red diamonds), matched to the original sample in number only, with the true sample plotted in blue.  The bootstrap modal values in each redshift bin and the derived evolution for this realisation are over-plotted in black.  \textbf{The bottom panel} (b) shows the distribution of measured gradients for 500 realisations where the AR galaxies are selected, with repeats, matching the number of objects in the actual sample redshift bins.  The results from three different scenarios are shown.  When the artificially redshifted galaxies are randomly sampled to only match the number of objects in each redshift bin is plotted as the grey shaded histogram.  The red histogram displays the results when also matching the samples by luminosity.  The actual measured evolution derived from the bright luminosity bin is plotted as the vertical line, with the associated errors plotted as dashed lines.}
  \label{fig:null_hypothesis_sizes}
\end{figure}

The results from this analysis are shown in Fig. \ref{fig:null_hypothesis_sizes}, where panel (a) shows one example of deriving the evolution from the AR sample, and panel (b) shows the distribution of derived gradients for 500 realisations (grey histogram).  The histogram of derived gradients is not centered on zero because the two highest redshift bins tend to have their typical sizes underestimated.  This is due to two factors, the sampling of the underlying population and the underestimation of the sizes of the largest galaxies.  With such small sample sizes in these two bins, the estimation of the mode is dominated by small number statistics.  The galaxies with sizes larger than the true typical size will have systematically underestimated sizes, however, hence biasing the typical size measurements to lower values.  The lack of bias from the simulations (Section \ref{fig:typicalSizeBias}) relies on sufficient sampling of the underlying population so that the mode is well-defined.  The gradient and errors derived from the bright luminosity bin are overplotted, showing that the derived size evolution in the bright bin is consistent with no size evolution in the underlying population.

One would ideally perform this test by matching the objects in both field and luminosity so that all luminosity evolution and surface-brightness effects are taken into account.  However, the samples are not large enough to simultaneously match between both those parameters.  We therefore also plot the histograms produced when matching in luminosity (each object in the sample is matched to an object from the AR sample within $M_{1500}\pm0.2$) and find very little difference to the derived size evolution from the AR sample.

It is possible that taking the derived size evolution from the mode estimator could mask any evolution in the spread of the distribution, which the median or the mean might be sensitive to.  Although it would not be suitable for comparing to predictions from theory we repeat the above analysis using the median and the mean of the size distribution.  We find in each case that we cannot reject the null hypothesis of no size evolution.  This test does not reject the possibility that there is evolution in the build up in the tail to large sizes (see Fig. \ref{fig:size_logNorm_allRedshifts}), or in the width of the lognormal size distribution, only that there is no firm evidence to support that scenario.

\subsection{The population of disturbed galaxies}

\subsubsection{Disturbed galaxies in the tail of the lognormal size distribution}

Previous studies suggested that the evolution in the measured size evolution is driven by the build up in number of galaxies contributing to the tail to large sizes.
Although tracking the typical sizes of high-redshift galaxies requires us to plot the peak of the lognormal distribution (see Fig. \ref{fig:size_logNorm_allRedshifts}), rather than the mean size in real space, it is still instructive to investigate the nature of the galaxies contributing to the tail of the distribution at large sizes. If the objects in the tail were primarily disturbed, that would indicate that processes contributing to this disturbance could also be affecting the size measurement independently of the underlying halo spin parameter.  Possible mechanisms for disturbed morphologies are clumpy star forming regions in the underlying disc or mergers of distinct systems.  Either of these mechanisms would likely render the rest-frame UV unsuitable for studying the underlying mass profile of the galaxies and could affect the measured size distribution.

In this study the asymmetry measurement is used to characterise the morphological disturbance of galaxies.  More accurately it provides an indication of whether there are any features associated with the galaxy lying above the background noise that are inconsistent with the profile being described as smooth and relaxed (as discussed in Section \ref{section:asymmetryMeasurements} the measurements are sensitive to features in the central regions of these objects).  Considering that measurements taken from the rest-frame UV trace the star-forming regions in the galaxy, a large asymmetry value does not necessarily indicate a disturbed mass profile.

\begin{figure}
  \centering
   \subfigure{\includegraphics[width=3in]{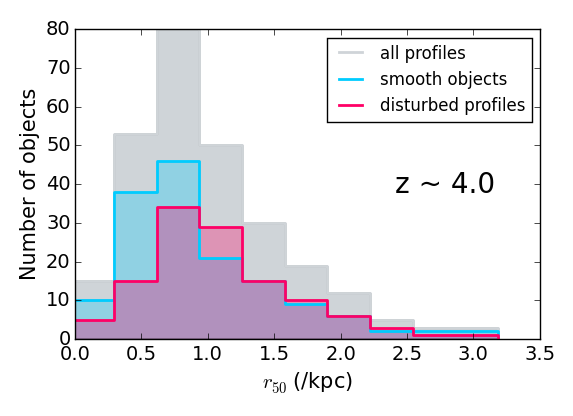}} 
   \subfigure{\includegraphics[width=3in]{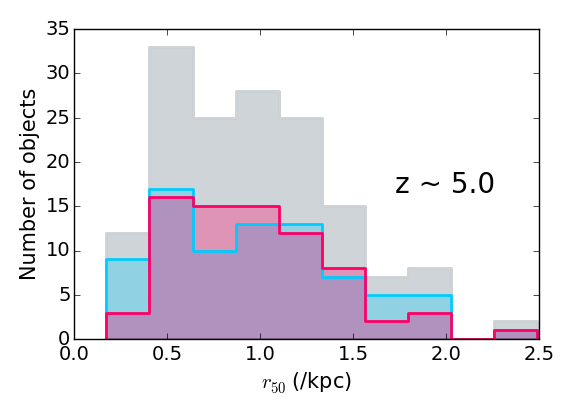}} 
   \subfigure{\includegraphics[width=3in]{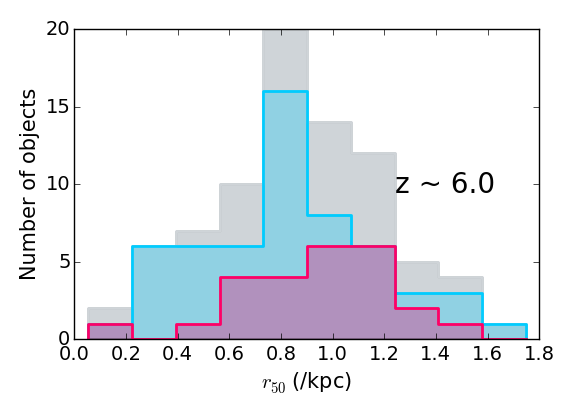}} 
  \caption{The measured size distribution in three redshift bins, $z\sim4$, 5 and 6.  The sizes for all objects in the redshift bin are plotted as the grey shaded region.  The distribution is then plotted separately for objects with asymmetry values suggesting disturbed profiles (purple shading) and \textbf{undisturbed} profiles (light blue shading).}
  \label{fig:size_log_normal}
\end{figure}

In Fig.~\ref{fig:size_log_normal} the object size distributions are plotted in three different redshift bins, $z\sim4$, 5 and 6.  The distribution is plotted for a constant luminosity range, $-21 <\,$\MUV$\,< -20$  within which each object has reliable asymmetry measurements. The size distributions are plotted for the whole redshift bin, as well as for the disturbed and smooth profiles separately.

This figure shows that the objects with disturbed morphologies are not all large, they have a fairly uniform distribution of sizes. In fact, a two-sample Kolmogorov-Smirnov (KS) test performed on each of the samples for galaxies with $r_{50}>1$ kpc gives p-values of 0.8, 0.4 and 0.4 for redshift bins $z\sim4$, 5 and 6 respectively.  Therefore we cannot reject the null hypothesis that the distribution of smooth and disturbed galaxies with sizes $>1$ kpc are drawn from the same sample.  When performing the KS test without a minimum size constraint we find p-values of 0.001, 0.8 and 0.3 respectively, suggesting that at redshift $z\sim4$ there are more smooth profiles with small sizes.  However, given that the smaller galaxies are less likely to be measured as disturbed we cannot dis-entangle this from the inherent resolution  constraints.

\subsubsection{Evolution in fraction of disturbed galaxies}
\label{section:morphologyEvolution}

\begin{figure}
  \centering
   \includegraphics[width=3.2in]{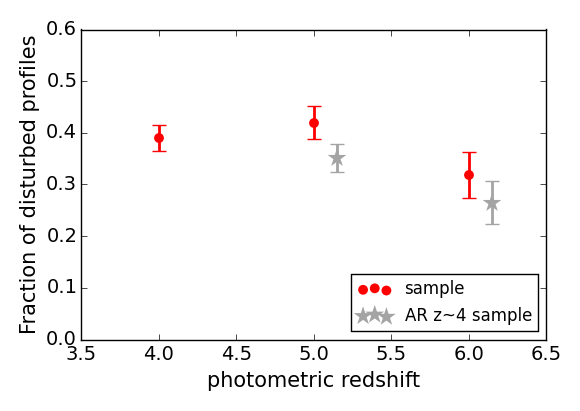}
  \caption{The fraction of disturbed profiles with \MUV$\,<-20$ plotted as a function of redshift.  Disturbed profiles are defined as those galaxies with asymmetry measurements that differ from the distribution of values obtained for synthetic, symmetric single S\'ersic profiles matched in luminosity.   The red circles represent the measurements made from the actual sample and the grey stars represent the measurements made from AR galaxies taken from the $z\sim4$ sample and matched to the original sample in size and UV luminosity (see text for details).  The errors plotted for the actual sample show the standard deviation for a binomial distribution if there is an even probability that the profile is smooth or disturbed, and so display the uncertainty in the measured fractions due to the number of galaxies in each redshift bin.  The errors for the UV and size-matched AR sample are derived from randomising the selection of matched objects and re-sampling with repeats.}
  \label{fig:fraction_asymmetric}
\end{figure}
The quantifiable morphology measurements made with this method allow comparison of morphologies from images of very different depths and PSFs.  This allows us to search for any evidence of evolution in the fraction of objects that are measured as disturbed while taking account of all observational biases in a self-consistent way.

To look for any evidence of evolution in the incidence of morphological disturbance amongst galaxies with redshift, the fraction of disturbed profiles with \MUV$\,<-20$ is plotted as a function of redshift in Fig.~\ref{fig:fraction_asymmetric}.  

The measured asymmetry is extremely sensitive to resolution and surface-brightness limits in the following ways:
\begin{itemize}
  \item The distribution of asymmetry values measured for symmetric, smooth profiles broadens significantly with decreasing flux, due to increased noise.
  \item The pixel scale and PSF broadening provide a resolution limit.  Features of disturbed profiles on small scales cannot be distinguished.  The size distribution of galaxies can then affect the measured fraction of disturbed profiles.
\end{itemize}
We therefore do not hope to provide an absolute fraction of disturbed profiles among the high-redshift galaxy population but instead look for trends in the fraction of most disturbed profiles with redshift.

Objects at different redshifts have their asymmetries measured in images of differing depths and resolutions.  These effects will potentially dominate any observed trend in disturbed fraction with redshift.
Therefore the measured fraction of disturbed profiles measured from the artificially redshifted $z\sim4$ sample are plotted (star symbols).  

The results of this analysis are presented in Fig.~\ref{fig:fraction_asymmetric}.  The asymmetry measurements are highly sensitive to size and surface brightness so that the distributions of size and UV luminosity are matched between the true galaxy sample and the AR galaxies.  Each galaxy in the true sample is assigned matches within the AR samples within $r_{50} \pm0.25~\textrm{kpc}$, \MUV$\,\pm0.25$.  A sample is then randomly drawn from these matches, with repeats.  The mean and standard deviation for the measured fraction is then plotted as grey points in Fig. \ref{fig:fraction_asymmetric}.  There are too few objects in the AR sample at $z\sim7$ and 8 to be able to match the distributions in both size and luminosity, and so we only investigate the fraction of disturbed profiles at $z<7$.

These results show that a galaxy at $z\sim6$ is just as likely to show a disturbed profile as one at $z\sim4$ with the same physical size and UV luminosity, at least to the extent that can be determined with current imaging depths and resolution.  These results are also qualitatively reproduced with a 95\% and 90\% cut to the asymmetry measure.  Here we provide the caveat that the asymmetry measurement used to determine whether or not an object is disturbed is not sensitive to features in the outer regions of the galaxies.  These results indicate that there is no evidence \textit{yet} for an increase in the fraction of objects showing very clumpy features (or possible multiple components) with redshift that might be indicative of the mechanisms of star formation in these galaxies.

\begin{figure*}
  \centering
  \includegraphics[width=7in, trim=5 0 5 0 , clip]{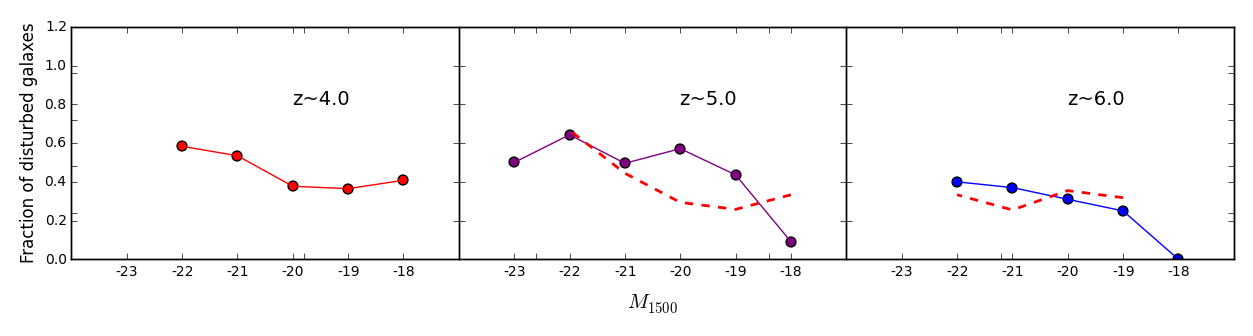}
   \caption{The fraction of disturbed galaxies as a function of UV luminosity for the three lowest redshift bins.  Only bins containing more than one object are plotted.  The red dashed lines in the middle and right-hand panels show the results for the artificially redshifted $z\sim4$ sample.}
     \label{fig:fraction_asymmetric_MUV}
\end{figure*}

In Fig. \ref{fig:fraction_asymmetric_MUV} we show the fraction of disturbed galaxies as a function of UV luminosity in the three lowest redshift bins.  The numbers of objects are too low in the two highest bins to be able to sub-divide the samples as a function of UV luminosity.   At each redshift we find that the fraction of disturbed profiles is fairly flat at $\sim0.4-0.5$, except at the faintest luminosities where surface-brightness limits make the measurement of disturbed features unfeasible. We also show the results for the full artificially redshifted $z\sim4$ sample.  As for the results shown in Fig.\ref{fig:fraction_asymmetric}, there is no significant difference between the slow decline in fraction of disturbed profiles with redshift and that measured from the AR sample.

\section{Discussion}

\subsection{How different size measurement techniques affect the measurement of the size-distribution of galaxies}

In this study we use a non-parametric CoG measurement to investigate the sizes of high-redshift galaxies.  Other studies have used alternative methods for estimating galaxy sizes, such as {\scshape SExtractor} (e.g. \citealt{Ferguson2004}; \citealt{Bouwens2004a}; \citealt{Oesch2010a})    and Galfit (e.g. \citealt{Ono2013}; \citealt{Shibuya2015}).  It is therefore important to discuss the strengths and weaknesses in all of these approaches and how they affect the measurement of the underlying size distribution of high-redshift galaxies.

{\scshape SExtractor} estimates the half-light radius with a method that is similar to the curve-of-growth algorithm used in this study.  Both algorithms use a non-parametric approach, measuring the increase in flux in successively larger circular apertures centred on the object.  Where they differ significantly is in the total flux estimate which is vital for determining the half-light radius ($r_{50}$).  {\scshape SExtractor} defines the total flux as that within a multiple of the first moment of the light distribution, $kr_1$, where $r_1$ is defined using an elliptical extension to the Kron `first moment' algorithm \citep{Kron1980}, and where $k$ is user-defined.  This first moment is defined from pixels meeting a user-defined S/N ratio.  For fainter galaxies, the total flux, and hence half-light radius becomes underestimated, as shown in the simulations in Section \ref{section:simulations}.

The CoG algorithm employed here uses an aperture of a fixed-size to measure the total flux.  This aperture ensures that any high surface brightness components within a lower surface-brightness object do not dominate the derived physical properties.  This is not the case when using pixel S/N to weight a pixel-by-pixel fit (as with {\scshape galfit}) or to a Kron-based measurement of an object's total flux (as with {\scshape SExtractor}).  This choice also allows for a clean selection mechanism based on measured total fluxes that is not dependent on an object's central surface-brightness which could otherwise bias the measured size-distribution.

However, although offering several advantages, our own simulations have shown that our CoG method will truncate the sizes of the largest galaxies, something which would not happen with {\scshape SExtractor}-based $r_{50}$ measurements. To overcome this bias we chose the size of the total flux aperture as a balance between an aperture large enough that we can still measure the size evolution of galaxies but small enough that we select large enough samples of galaxies in each redshift bin.  Our simulations show that our strategy does not affect the typical size measurement at $z\sim4$, and, if the typical galaxy size is evolving as rapidly as indicated by previous studies ($\propto(1+z)^{1-1.5}$), this bias would not affect the measurement of the evolution in typical galaxy sizes.  However, given that our results suggest a slower evolution of size with redshift, it is likely that our size estimates at $z=7-8$ will be underestimated.

Both {\scshape SExtractor} and the CoG algorithm employed here require PSF corrections to be made a postoriori.  To first approximation these corrections are applied in quadrature using a Gaussian approximation for the central portion of the HST WFC3 PSF.  These corrections are dependent on galaxy structure as displayed in Section \ref{section:psf}.  They will act to distort the size-distribution at very small sizes ($r_{50} < 0.5$ pixels), due to the non-Gaussianity of the central portion of the HST PSF.  We also show, however, that due to the extended wings in the HST PSF we require an additional constant size correction term (Fig. \ref{fig:psfCalibration}).  The magnitude of this correction is dependent of the size of the total flux aperture and the internal structure of the galaxy.  Using a fixed-size total flux aperture for the CoG algorithm allows for this correction to be applied optimally for objects with S\'ersic index $n=1$.  Given that this correction is constant for a fixed-sized total flux aperture, it will not significantly distort the measured size distribution when using the CoG measurement. Only if the underlying galaxy properties show a wide range of structural parameters will the correction affect the measurement of the underlying size distribution, but our simulations show that the change in the correction term for different S\'erisc indices is a small fraction of the required correction and so is a secondary effect.  Measurements obtained using {\scshape SExtractor}, however, have total flux apertures that are dependent on an object's properties and hence the correction would have to be calculated on an object-by-object basis and failure to provide this correction would bias the shape of the derived size distribution.

The other algorithm often used to measure the sizes of high-redshift galaxies is {\scshape galfit}.  It preforms parameterised surface-brightness fitting that can fit arbitrarily complex parameterised profiles to pixel data using $\chi^2$ minimisation.  The PSF is convolved with the parameterised profile before fitting to the image and so the PSF correction is clearly more accurate than that derived for the CoG and {\scshape SExtractor}-based sizes, especially for the smallest profiles.  Additionally, as with {\scshape SExtractor}, the outer parts of large galaxies are used in the fit and hence the tail to large sizes in the size distribution will be better reproduced using {\scshape galfit}.  However, it is worth stressing that to achieve this improved PSF correction and fitting to the low surface-brightness outskirts, one must resort to a parameterised measurement which has the following disadvantages.

At high-redshift, surface-brightness constraints mean that objects do not have high enough S/N to sufficiently constrain multiple components (e.g. bulge/disc, axis ratio, S\'ersic index and size).  For this reason, most studies using {\scshape galfit} resort to fixing the S\'ersic index at a given value.  The size-distribution will thus be distorted (as for CoG-derived sizes) if galaxies do show a range of galaxy structural parameters, or indeed if their underlying profiles are not well described by single S\'ersic profiles.  Additionally, the procedure uses a gradient minimisation algorithm to search for the minimum $\chi^2$ solution.  These methods are fast but do not sufficiently sample the parameter space, sometimes providing results for a local $\chi^2$ minima and unrealistic parameter uncertainties.  One way to address this is to run {\scshape galfit} over a grid of various parameter values (including background, S\'ersic index and axis ratio) in order to investigate the uncertainties using $\Delta\chi^2$ values (as performed in \citealt{McLure2013}), however this has not been implemented in current studies of the sizes of high-redshift galaxies using {\scshape galfit} to date.  Even addressing the uncertainties in the fits more rigorously does not take account of the possibility of complex profiles with clumps of star formation or asymmetries that can subsequently bias the fits. 

In summary, although {\scshape galfit} has several advantages, it does fundamentally rely on the assumption of smooth axisymmetric profiles which are not typically seen in the rest-frame UV, and can be potentially biased by high surface brightness features such as clumps.  The beauty of the non-parametric CoG algorithm is that it makes no assumptions about the underlying galaxy surface-brightness profile and is robust against high-surface brightness features dominating the total flux/size measurements.

\subsection{Comparison to previous work}

The size evolution for galaxies from $z\sim4$ to $z\sim8$ presented in section \ref{section:size_evolution} is shallower than that reported in many previous studies.  There are two main reasons for this discrepancy, the measurement of \textit{typical} galaxy size at each redshift and the redshift baseline over which the measurements are made.  The bright flux limits imposed by the simulations (Table \ref{table:fluxLimits}) are also more strict than employed in previous studies and are based on measurements from within large apertures.  Although the inclusion of fainter objects may bias typical {\scshape SExtractor}-based size estimates to smaller sizes ({\scshape SExtractor} systematically underestimates sizes of the faintest galaxies, see Section \ref{section:measurementCalibration}), mean estimates of typical sizes from our sample agree well with previous works, suggesting that this particular effect is likely to be small. 

The studies by  \cite{Ferguson2004} and \cite{Bouwens2004a} find size evolution consistent with that expected for objects selected at constant halo mass ($r_{50}\propto(1+z)^{-1}$).  These studies fit to sizes over a wide range of redshifts, from $1\lesssim z\lesssim5$ in the case of \cite{Ferguson2004} and from $2\lesssim z\lesssim 6$ in the case of \cite{Bouwens2004a}.  Subsequent studies have extended the coverage out to higher redshifts; \cite{Oesch2010a} add $z\sim8$ selected galaxies; \cite{Ono2013} add $z\sim9$ galaxies plus robust size estimates from deeper imaging of $z\sim7-8$ galaxies.  Although \cite{Oesch2010a} quote slightly steeper evolution than the earlier studies  (gradients of $n=-1.32 \pm 0.52$ and $n=-1.12\pm0.17$ in the faint and bright luminosity bins respectively) their results are still formally consistent with constant halo velocity evolution.  \cite{Ono2013} find slightly tighter constraints with a measured gradient of $n=-1.30\pm0.13$ suggesting that the evolution lies somewhere between the two scenarios of constant halo circular velocity or halo mass.
 
All of the studies mentioned above use a mean estimator to describe the typical sizes of galaxies at each redshift.  The build up of the tail to large sizes in the size-distribution to low redshift would naturally steepen the fit to the size evolution compared to a modal estimate (see Fig. \ref{fig:size_logNorm_allRedshifts}). In fact we show that the modal estimate made at $z\sim4$ is significantly different to what would be measured for the mean in real space.  Taking this into account, however, steepens our derived evolution to $n\sim-0.9\pm0.3$, which is only marginally shallower than that of other studies, and consistent with evolution at constant halo mass.  The other possible reason for such a difference is in the redshift baseline used to constrain the measured evolution.  This study addresses only the evolution of sizes in galaxies at $z\gtrsim3.5$, below which U-band imaging would be required to provide consistent rest-frame size measurements.  For all redshifts studied here, the Universe was less than $\sim1.5$ Gyr old.  It is possible that including consistent measurements at lower redshifts could steepen the derived evolution.

\cite{Huang2013} study galaxies at $z\sim4-5$ and find a $\sim13\%$ evolution in size between these two redshifts from the \textit{peak} of the size distribution in each bin.  This corresponds to a gradient of $\sim-0.67$.   This is slightly steeper than that measured in the high luminosity bin ($n=-0.20\pm0.26$), although is in agreement to within $\simeq2\sigma$.    

A complementary study of disk growth, \cite{Fathi2012}, claims a factor of $\sim8$ size increase from $z\sim5.8$ to $\sim0$ for the brightest disc galaxies in their sample ($-24 <\,$\MUV$\,< -22$), with most of the evolution occurring between $z\sim2$ and 5.8.  This would suggest much faster evolution than constant halo mass. \cite{Fathi2012} do not claim such fast evolution for fainter galaxies more comparable to the sample presented here, primarily due to spectroscopic incompleteness at the highest redshifts. The sizes reported in their Table 1 suggest shallower evolution for $-20<\,$\MUV$\,<-22$ galaxies.  It is worth noting that this study also measures disc scalelengths from the same observed filter and is prone to uncertainties in the morphological k-correction applied to the highest redshift galaxies. 

Another factor that possibly contributes to the different measured size evolution is the treatment of multiple component systems.  In this study, multiple component systems are treated as a single object.  It is possible that they are multiple star forming clumps in an underlying system, but it is also possible that some, or all of them are instead separate systems that are close to each other and hence should be treated separately.  However, using the asymmetry measure to remove any objects with `disturbed' morphologies from the analysis does not significantly affect the results. 

\subsubsection{Comparison to Shibuya et al. (2015)}
\label{section:shibuyaComparison}

\begin{figure}
  \centering
   \subfigure{\includegraphics[height=2.35in]{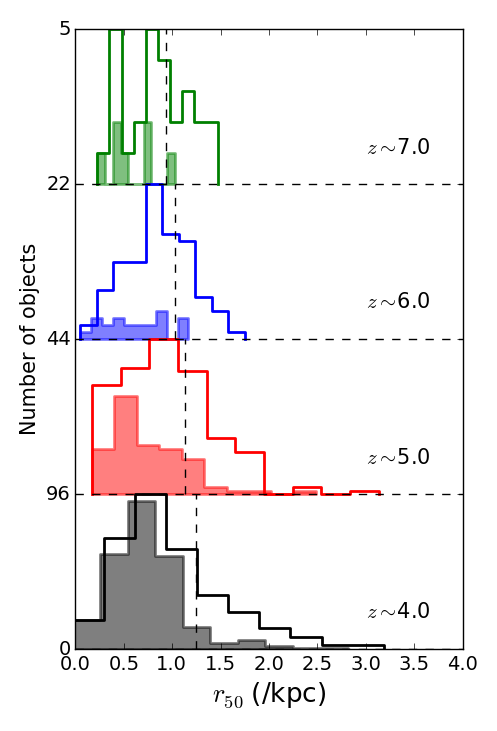}}
  \subfigure{\includegraphics[height=2.35in]{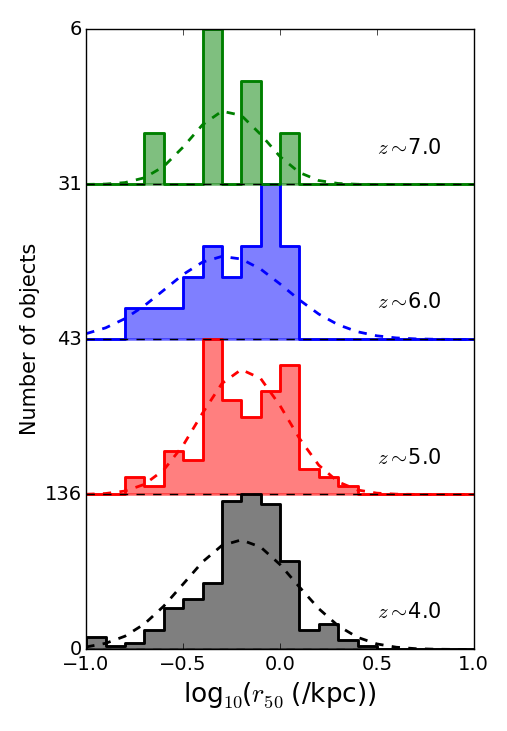}}
  \caption{\textbf{Left:} The derived size distributions for our sample (thick lines) and again for all objects satisfying a selection mechanism mimicking that employed in S15 (shaded histograms, see text for details).  \textbf{Right: }The logarithm of the size distributions for the objects passing the Shibuya et al. (2015) selection.  Dashed lines show the best-fitting normal distributions, the peaks of which are plotted as dark blue filled circles in Fig. \protect\ref{fig:shibuyaComparison2}.}
  \label{fig:shibuyaComparison1}
\end{figure}
\begin{figure*}
  \centering
  \includegraphics[width=5in]{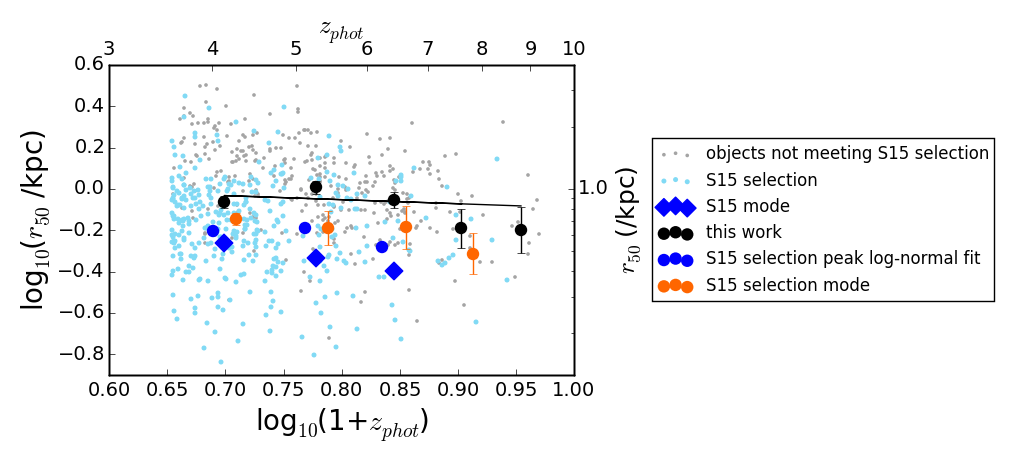}
  \caption{A comparison of results using our selection criteria and the selection criteria employed by \protect\cite{Shibuya2015} (S15, S/N$>15$ with in a 0.35\asec diameter aperture).  The small points show the objects in our sample.  Those coloured light blue would also be selected by the S15 selection function.  We display both the results from S15 (blue diamonds, taken from their modal values from fitting a lognormal distribution to the measured sizes) and the results from our measurements when employing the S15 selection.  The orange circles show the bootstrapped mode estimated from the S15 selection applied to our sample and the blue circles show the measured peaks derived from the best-fitting normal distribution to the logarithm of the size-distribution (see Fig. \ref{fig:shibuyaComparison1}).  Small offsets in the $x$-direction are added for clarity.}
  \label{fig:shibuyaComparison2}
\end{figure*}

The recent work by \cite{Shibuya2015} (S15) investigates the difference in derived evolution when employing different statistics to measure the typical galaxy sizes.  As such, this work is the most directly comparable to our own, and yet they find that galaxy sizes continue to follow evolution steeper than $r_{50}\propto(1+z)^{-1}$ to high redshifts, despite the statistic employed (including measurements of the peak in the distribution).

Within the redshift range overlapping with our sample, their selection criteria is based on colour criteria.  The objects are then subjected to a secondary cut to ensure that reliable size estimates can be obtained.  This secondary cut requires the object to have S/N$>15$ within an aperture of 0.35\asec diameter (Shibuya, private communication).  We have already discussed in Section \ref{section:sizeBias} the importance of the selection function for measuring the size-distribution of galaxies robustly.  Here we investigate the impact of this secondary cut on our derived typical galaxy sizes by subjecting our sample to the same selection criteria.  Fig. \ref{fig:shibuyaComparison1} displays our measured size distributions before and after applying the S15 selection.  The new size distributions (filled histograms) are clearly more skewed to smaller sizes than the original distributions.  

We can then compare the derived typical sizes in Fig. \ref{fig:shibuyaComparison2}.  The modal values derived using our bootstrapping method are smaller than those measured from our full sample, although still larger than the modal values reported in S15.  The S15 results are derived from fitting a lognormal distribution to their recovered size distributions.  We mimic these measurements by fitting normal distributions to the logarithm of the sizes (Fig. \ref{fig:shibuyaComparison2}, right-hand panel).  The peak of the best-fitting distributions for the $z\sim4,5$ and 6 bins are plotted in Fig. \ref{fig:shibuyaComparison1} and show much better agreement with the S15 results. 

The discrepancy in derived sizes between this work and that of \cite{Shibuya2015} is therefore dominated by the selection mechanism employed in that work, which preferentially adds smaller galaxies to the sample and so skews the measured size distributions to smaller sizes .  Additionally their method of fitting a lognormal function to their measured size distribution can further skew the typical size measurements to smaller sizes if the tail to large sizes is not accurately reproduced by the selection mechanism.

\subsection{Bright flux cuts in the presence of a size-luminosity relation}

\begin{figure}
  \centering
   \subfigure{\includegraphics[width=3in]{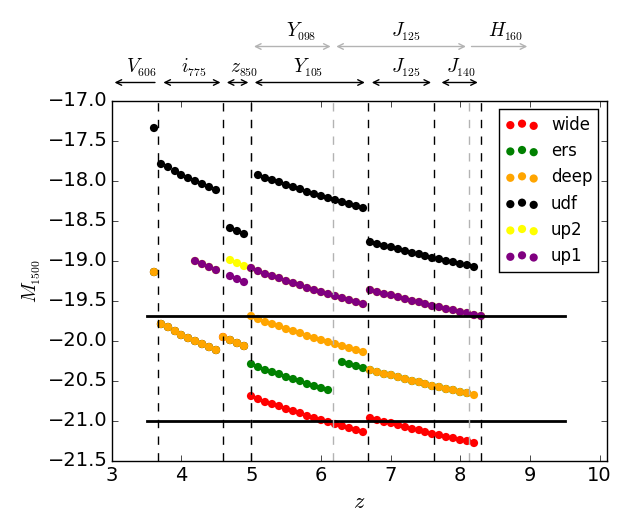}}
  \subfigure{\includegraphics[width=3in]{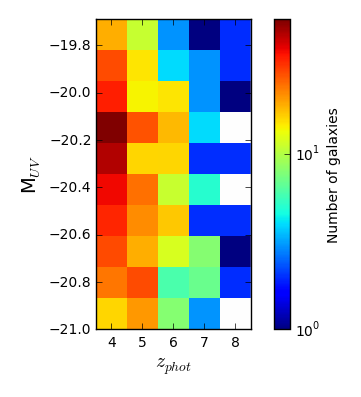}}
  \caption{\textbf{The top panel} shows the distribution of flux cuts in absolute magnitude as a function of redshift.  The flux cuts applied to the samples are displayed relative to the limits of the bright luminosity bin (horizontal black lines).  The limits in each field are colour coded according to the legend.  Different images are used to select objects at different redshifts (as shown with the labels above the top axis and the dashed vertical lines), hence the irregular magnitude cuts as a function of redshift for each field.  The deep, ERS and wide data all have the same limits in the optical filters and so only the flux cuts for the deep field are shown.  This is also true for the deep and ERS $J_{125}$ image cuts and where limits for the two parallel fields overlap, only the limits for up1 are shown.  \textbf{The bottom panel} shows the number of objects contributing to sub-bins in the bright luminosity bin across the whole redshift range (white squares mean no objects contribute to that bin).  These numbers are used as inverse weights for objects occupying the bins when assessing whether the presence of an underlying strong size-luminosity relation could be biasing the derived size evolution.} 
  \label{fig:fluxCutsAbsMag}
\end{figure}

Although our selection mechanism is designed to prevent the preferential introduction of small galaxies at faint fluxes into our sample, it is still possible that the derived sizes could be biased by the preferential introduction of objects at bright luminosities from shallower images.  The inhomogeneity of the selection function is displayed in Fig. \ref{fig:fluxCutsAbsMag}.  Some of the fields only contribute objects at the bright end of the luminosity bin and this has potential to bias the derived sizes in the presence of a strong size-luminosity relation.    The size-luminosity relation that we measure is not incredibly steep, in agreement with the results of \cite{Huang2013} and Shibuya et al. (2015) but we can investigate the importance of potential biases using a weighted median measure.

\begin{figure}
  \centering
  \subfigure{\includegraphics[width=3in]{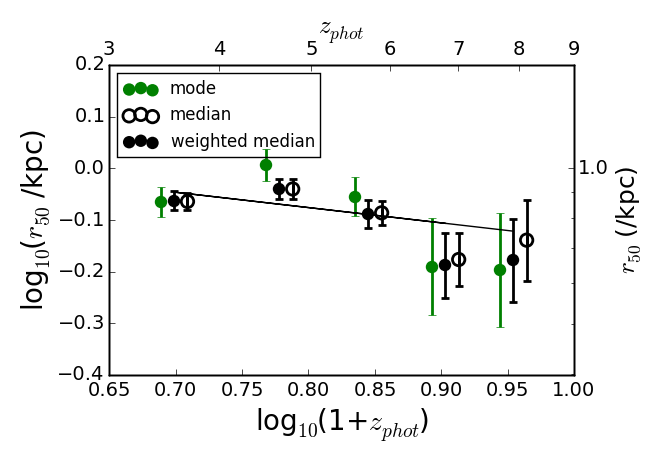}}
  \subfigure{\includegraphics[width=2.7in]{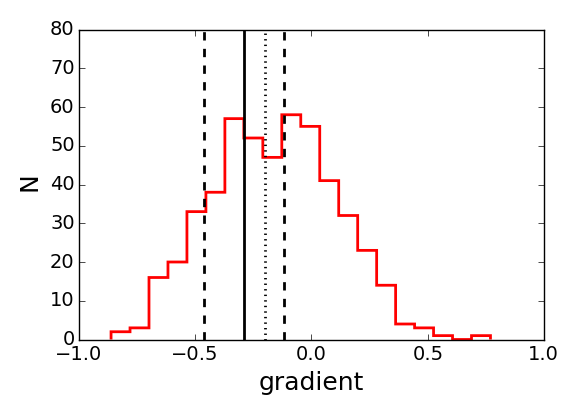}}
  \caption{\textbf{The top panel} shows the typical size evolution for the bright luminosity bin as measured from modal (the same as the points shown in Fig. \ref{fig:size_redshift_bright}), median and weighted median estimator as indicated in the legend.  The estimators have been offset artificially in the $x$-direction for clarity.  Errors are estimated from bootstrapping as for the original modal analysis.  The weights employed for the weighted median estimates use the inverse number of objects within sub-bins in luminosity as shown in Fig. \ref{fig:fluxCutsAbsMag}, bottom panel.  The measured evolution using the weighted median estimator is $n=-0.29\pm0.17$ and is not significantly different to that measured with the median estimator without weighting. \textbf{ The bottom panel }shows the associated null test with the median estimator and luminosity matching between actual sample and AR sample.  This clearly demonstrates that when we measure the evolution with the median, we still cannot reject the null hypothesis of no size evolution, with the gradient and associated error measured from median estimator plotted in solid black and dashed lines respectively.  The gradient measured for the modal estimates is plotted as the black dotted line.} 
  \label{fig:weightedSizeEvolution}
\end{figure}

We weight the objects to ensure even contribution according to UV-luminosity across the entire bright bin.  We do this by splitting each redshift bin into 10 further bins according to UV-luminosity and then a single objects' weight is set by the inverse number of objects contributing to that luminosity bin.  This weighting ensures that any evolution in the underlying luminosity function does not affect the measured size evolution by contributing proportionally more objects at the faint end of the bin at high redshifts.  It also ensures that any excess of bright objects added to the bin in fields with bright corresponding flux cuts does not bias the measured sizes.  For these size measurements we use a weighted median estimator rather than the mode and the results are displayed in Fig. \ref{fig:weightedSizeEvolution}.  We also show, in the bottom panel the corresponding null hypothesis test using the median estimator for consistency.

The results show that there is very little change when applying the UV-luminosity based weighting to the results.  This is mostly explained by the fact that the distribution of UV-luminosity is, on the most part, flat in each redshift bin, with the exception of the $z\sim4$ bin.  Additionally, the measured size-luminosity relation is not very steep. An object at the very faint end of the luminosity bin is only expected to be, on average, 80\% smaller than an object at the very bright end of the bin (with slope of $\beta\sim0.2$ for the size-luminosity relation).

\subsection{On the lack of evolution in the incidence of morphological disturbance}
\label{section:discussionClumpy}

In Section \ref{section:morphologyEvolution} we present our results of investigating the evolution in the incidence of morphological disturbance in bright (\MUV $<-20$) galaxies between $4<z<6$.  We compare our measured fraction of disturbed galaxies to that derived from the artificially redshifted $z\sim4$ sample and find no significant evolution.  From our results we find no strong evidence that the incidence of objects showing large-scale disturbances, either due to asymmetries in a smooth profile or due to distinct clumps, is increasing between these redshifts.

One likely cause of objects being measured to be `disturbed' is due to clumpy star formation.  It is not the only possibility but visual inspection shows that it is prevalent at high redshifts (e.g. see Fig. \ref{fig:postageStamps}) and mergers are less likely to show strong disturbance unless the objects are still distinctly separated (i.e. also resembling a clumpy morphology), due to the insensitivity of the asymmetry measurement to low surface-brightness features in the outskirts of galaxies (discussed in Section \ref{section:asymmetryMeasurements}).  Therefore, to determine what physical features this measurement is sensitive to, we performed simulations in which we added single clumps to smooth single S\'ersic profiles to investigate what clump properties change the asymmetry measurement from smooth to disturbed.

The evolution in the measurement of the disturbed fraction is sensitive to different scales (due to changing angular diameter distance and measurement image resolution) and SFR surface densities in each redshift bin.  As we are always comparing to the artificially redshifted $z\sim4$ sample, our sensitivity to evolution over the range $4<z<6$ is primarily driven by that data at $z\sim6$. The results are also primarily driven by the measurements from the GOODS-S deep image because $\sim80$\% of the galaxies at $z\sim5-6$ with M$_{1500}<-20$ are selected from this field.   To understand what features the measurement is sensitive to, it is therefore sufficient to consider the features that provide a `disturbed' measure in the GOODS-S deep $Y_{105}$ image, which captures the rest-frame 1500\AA~emission at $z\sim6$. 

From these simulations we find that the asymmetry measurement is essentially sensitive to features with peak fluxes separated by at least 3 pixels, or $\sim1$kpc at $z\sim6$ (the core of the PSF has a FWHM that corresponds to $\sim0.56$ kpc).  For individual (unresolved) components the $3\sigma$ SFR limit of this image is $\sim5.6$\Msun/yr.  However, our simulations find that un-resolved clumps with total SFR $\gtrsim17$ \Msun/yr, or fraction of flux in the clump compared to the total galaxy flux ( $f_c/f_g$) $>0.4$ result in P(A)$>0.98$ (for details, see Appendix \ref{Appendix:clumps}).  This measurement is only mildly dependent on underlying object morphology with	 discier profiles ($n<2.5$) showing disturbed measurements for fainter clumps than non-discy galaxies ($n>2.5$). 

The surface-brightness limits mean that the measurement is only sensitive to galaxies with clumps that contain a large fraction of the total flux (or star formation) of the galaxy and larger samples at depths comparable to HUDF are required to provide a measurement sensitive to typical clump SFRs ($\sim 2.7$ \Msun/yr at $z\sim3$ in the sample of \citealt{Livermore2015}).

We are therefore not going to see significant improvement in the determination of the prevalence of morphological disturbance within the galaxy population at high redshifts until the launch of the \textit{James Webb Space Telescope} (\textit{JWST}), which will provide a longer redshift baseline, extending this measure to higher redshifts, and lower surface-brightness limits, allowing the analysis of the disturbed fraction to lower clump SFR limits.

\subsection{The hazards of measuring sizes from the rest-frame UV}

 When we compare the observed size evolution to that expected for the underlying halo properties, we effectively end up asking whether this constant UV selection is closest to a constant halo mass or halo circular velocity selection.  Our current measurement of size evolution would lead us to state that it is shallower than that expected for either of these selections.  In fact, although we cannot reject the null hypothesis of no evolution at bright luminosities, we cannot reject the case of constant halo mass evolution at $\gtrsim3\sigma$ (see section \ref{section:nullHypothesis}), the uncertainties in the faint luminosity bin do, however, formally allow for constant halo mass evolution.   It is not clear whether the shallow measured evolution is due to disc galaxies not following the growth of their parent haloes at these redshifts, whether a constant UV selection is not suitable for studying disc growth as a function of underlying halo properties, or whether galaxies at these epochs are not, in fact, steadily growing, relaxed discs.

If we want to compare the measured size evolution to halo properties in order to provide constraints for galaxy formation models, we first need to consider whether the size measured from the rest-frame UV would sufficiently trace the size of an underlying disc.  The star formation tends to trace gas density rather than stellar mass, so for relaxed systems with a gas density profile that \textit{does} trace stellar mass profile, it would be reasonable to expect that the measured size evolution is representative of that of an underlying disc.  If, however, there are modes of star formation that distribute the gas throughout the disc (such as mergers or disk instabilities induced by accretion of cold gas, e.g. those expected to produce the clump-cluster galaxies presented by \citealt{Elmegreen2005}), then the UV luminosity and inferred size will differ significantly to that of a relaxed system of the same size.

Our only means of assessing whether the rest-frame UV may be suitable for investigating the evolution in the mass profiles of galaxies at this time is via their rest-frame UV morphologies.  Our results would indicate that there is no measurable evolution in the most disturbed fraction of galaxies.  These galaxies are most likely to have rest-frame UV profiles that do not trace their underlying mass profiles (if due to clumpy star formation or mergers).  There is therefore no evidence that any measured size evolution is impacted by a change in prevalence of a certain mode of star-formation that can affect the UV luminosity and rest-frame UV size independently of the underlying halo properties.  Also those galaxies with the most disturbed profiles are not solely responsible for driving the apparent build-up of galaxies in the tail of the size distribution, i.e. there is no indication from current imaging constraints that any part of the size distribution of galaxies is predominantly occupied by disturbed objects.

Clearly, to resolve whether the lack of size evolution observed in this study is indicative of a lack of evolution in the physical sizes of high-redshift galaxies we need measurements of galaxy mass profiles.  For this we require high resolution, rest-frame optical/NIR imaging and to obtain this over the redshift range studied here we require imaging from \textit{JWST}.

\subsection{The validity of the relaxed disc assumption}

The second important question to address is whether these objects can provide constraints on disc growth i.e. are they all relaxed discs and are discs growing steadily with time? 

The current framework used to link observed galaxy size evolution to halo properties relies on a number of assumptions, including that the total angular momentum of the baryons is equal to a fixed fraction of the total angular momentum of the dark matter halo.  It does not take into account accretion of material that is not aligned with the initial angular momentum of the halo.  

The recent studies of \cite{Danovich2012,Danovich2014} investigate in more detail the angular momentum transfer on to disk galaxies with gas transport via streams, rather than wide angle cylindrical infall.  These results are based on haloes selected at $z\sim2.5$ that later evolve into massive elliptical galaxies.  They find that the direction of angular momentum of the disc is most closely correlated to the dominant stream \citep{Danovich2012}, but that the overall spin of the disk is likely to be only moderately smaller than that of the dark matter halo \citep{Danovich2014}.

However, other studies suggest that while a galaxy is initially forming, the relaxed disc state may be transitory  \citep{Sales2012,Padilla2014}.  \cite{Sales2012} even investigated the correspondence between morphological parameters and underlying halo parameters from a sample of 100 parent haloes with halo mass similar to that of the Milky Way in the GIMIC gasdynamic simulation. They report that the most important factor driving galaxy morphologies is not the underlying halo properties, but the \textquotedblleft coherent alignment of the angular momentum of baryons that accrete over time to form a galaxy\textquotedblright.

Dynamical measurements of the galaxies selected here are not available to us at this time. However, observations of a small sample of lower redshift (13 galaxies at $z\sim2.5$) star forming galaxies indicate that lower-mass objects with higher gas fractions seem to be dispersion dominated while larger, more massive systems have higher velocity shear \citep{Law2009}. 

It is beyond the scope of this paper to determine whether all the selected galaxies are rotationally supported discs, or even whether the disturbed morphologies indicate merger activity or disc instabilities.  We do show, however, that there is a range of different types of galaxies selected, and not all of them display evidence of smooth, axisymmetric profiles.  Although it is possible that warping of discs could produce high asymmetry measurements, the presence of distinct clumps in many of these systems indicates the likelihood that the star formation is not tracing the underlying mass distribution in the same way as for smooth profiles.  The other assumptions inherent in trying to compare galaxy properties to halo properties need to be tested, however, as the studies mentioned in this discussion suggest that the sample may also contain dispersion dominated systems that are not appropriate for comparing to a scenario for disc growth.

\section{Conclusions}

We have measured the evolution of galaxy sizes and incidence of morphological disturbance for a sample of galaxies in the redshift range $4<z<9$.  They have been selected using photometric redshifts from the CANDELS GOODS-S, HUDF and HUDF-parallel fields.  

The size measurements reported are half-light radii taken from images closest in wavelength to $\lambda_{rest}=1500$\AA ~using a non-parametric CoG.  We find that this measurement technique and {\scshape SExtractor} \textit{both} systematically underestimate the sizes of the largest galaxies.  Additionally, at faint magnitudes the \textit{typical} galaxy size (the peak in the a lognormal size distribution) becomes systematically underestimated due to decreasing completeness.  We find that a completeness cut as high as 50\% is required to prevent biasing the \textit{typical} size measurements by preferential detection of small systems. 

Image-dependent flux limits are set from simulations of single S\'ersic profiles to ensure that typical galaxy sizes are not underestimated.  Simulations with galaxy sizes distributed lognormally show that typical size measurement should not be significantly biased due to this effect when a modal size estimate is used and the size distribution is well sampled.

The PSF correction employed is based on simulations of $n = 1$ single S\'ersic profile using a correction in quadrature for the core of the PSF and a further constant correction due to the extended wings. Our simulations show that this correction can distort the size distribution at the smallest sizes ($r_{50} < 0.5$ pixels).

Measuring the size evolution of $4<z<9$ galaxies from the \textit{peak}  in the lognormal size distribution of the form $r_{50}\propto(1+z)^n$ we find $n=-0.20\pm0.26$ in the (0.3-1)L$_*$ luminosity bin and $n=-0.47\pm0.62$ in the (0.12-0.3)L$_*$ luminosity bin.  Although we cannot reject the case of evolution consistent with that expected for disc galaxies selected at constant halo mass, we note that these results are consistent with no evolution in galaxy sizes.  We set up a test for whether we can reject the null hypothesis of no size evolution in the bright luminosity bin by artificially redshifting the $z\sim4$ sample into each successive redshift bin and mimicking the sampling of the underlying population.  This test shows that we cannot reject the null hypothesis of no size evolution and furthermore, the weak measured evolution can be explained by under-sampling of the population in the highest redshift bins, making measurement of the mode too uncertain.

Measurement of the typical galaxy sizes from the peak in the lognormal size distribution as a function of luminosity gives a shallow measurement of the size-luminosity relation ($r_{50}\propto\textrm{L}^{0.14\pm0.06}$, weighted average of all redshift bins) that shows no evidence of evolution  across the redshift range probed.

Investigation of the biases in our technique show that the CoG based measurements artificially truncate the size distribution at large sizes.  We use simulated profiles with a lognormal size distribution to show that this would not significantly affect the measurement of \textit{typical} galaxy sizes at $z\sim4$ when using a modal estimator, and if the sizes of galaxies evolve $\propto(1+z)^{n}$ with $1 < n< 1.5$  the typical galaxy sizes in higher redshift bins will also remain un-biased.  We measure a shallower evolution, however, meaning that the typical sizes in the highest redshift bins are likely to be biased, affecting the measured size-luminosity relation at $z\sim7-8$.

To investigate any evidence for evolution in the incidence of disturbed morphologies, we use an image-dependent specifier for whether or not the galaxy can be measured as `disturbed'.  A single non-parametric measure for galaxy structure (the asymmetry) is compared to measurements from populations of simulated single S\'ersic profiles to see whether the galaxy measurements are significantly different to those measured from smooth axisymmetric profiles matched in size and flux.  The fraction of `disturbed' profiles is compared to that measured from galaxies drawn from the artificially redshifted $z\sim4$ sample, with matched distributions of size and UV luminosity.  We find no evidence for evolution in the `disturbed' fraction of galaxies, with a galaxy at $z\sim6$ having the same probability of being labelled `disturbed' as a galaxy at $z\sim4$ matched in luminosity and size given the current surface-brightness and resolution limits available at this time.

\section*{Acknowledgments}

ECL would like to acknowledge financial support from the UK Science and Technology Facilities Council (STFC) as well as the ERC via an Advanced Grant under grant agreement no. 321323-NEOGAL.  RJM acknowledges the support of the European Research Council via the award of a Consolidator Grant (PI McLure). JSD acknowledges the support of the European Research Council via the award of an Advanced Grant, and the contribution of EC FP7 SPACE project ASTRODEEP (Ref. No.: 312725). ABR acknowledges the award of STFC PhD studentships.  AD acknowledges support from ISF grant 24/12, NSF grant AST-1010033,
and I-CORE Program of the PBC and ISF grant 1829/12. This work is based on observations taken by the CANDELS Multi-Cycle Treasury Program with the NASA/ESA HST, which is operated by the Association of Universities for Research in Astronomy, Inc., under NASA contract NAS5-26555.  This research has benefitted from the SpeX Prism Spectral Libraries, maintained by Adam Burgasser at http://pono.ucsd.edu/~adam/browndwarfs/spexprism.

\bibliographystyle{mn2e}
\bibliography{library}

\begin{thebibliography}{50}
\expandafter\ifx\csname natexlab\endcsname\relax\def\natexlab#1{#1}\fi

\bibitem[{Barnes \& Efstathiou(1987)}]{Barnes1987}
Barnes J., Efstathiou G., 1987, ApJ, 319, 575

\bibitem[{Beckwith {et~al}\mbox{.}(2006)Beckwith, Stiavelli, Koekemoer,
  Caldwell, Ferguson, Hook, Lucas, Bergeron, Corbin, Jogee, Panagia, Robberto,
  Royle, Somerville, \& Sosey}]{Beckwith2006}
Beckwith S. V.~W. {et~al.}, 2006, AJ, 132, 1729

\bibitem[{Bertin \& Arnouts(1996)}]{Bertin1996}
Bertin E., Arnouts S., 1996, A{\&}AS, 117, 393

\bibitem[{{Bertin, E.} \& Arnouts(1996)}]{BertinE.1996}
{Bertin, E.}, Arnouts S., 1996, A\&A Supplement

\bibitem[{Bouwens {et~al}\mbox{.}(2004{\natexlab{a}})Bouwens, Illingworth,
  Blakeslee, Broadhurst, \& Franx}]{Bouwens2004a}
Bouwens R.~J., Illingworth G.~D., Blakeslee J.~P., Broadhurst T.~J., Franx M.,
  2004{\natexlab{a}}, ApJ, 611, L1

\bibitem[{Bouwens {et~al}\mbox{.}(2007)Bouwens, Illingworth, Franx, \&
  Ford}]{Bouwens2007}
Bouwens R.~J., Illingworth G.~D., Franx M., Ford H., 2007, ApJ, 670, 928

\bibitem[{Bouwens {et~al}\mbox{.}(2011)Bouwens, Illingworth, Oesch,
  Labb{\'{e}}, Trenti, van Dokkum, Franx, Stiavelli, Carollo, Magee, \&
  Gonzalez}]{Bouwens2011}
Bouwens R.~J. {et~al.}, 2011, ApJ, 737, 90

\bibitem[{Bouwens {et~al}\mbox{.}(2004{\natexlab{b}})Bouwens, Illingworth,
  Thompson, Blakeslee, Dickinson, Broadhurst, Eisenstein, Fan, Franx, Meurer,
  \& van Dokkum}]{Bouwens2004}
Bouwens R.~J. {et~al.}, 2004{\natexlab{b}}, ApJ, 606, L25

\bibitem[{Bullock {et~al}\mbox{.}(2001)Bullock, Kolatt, Sigad, Somerville,
  Kravtsov, Klypin, Primack, \& Dekel}]{Bullock2001}
Bullock J.~S., Kolatt T.~S., Sigad Y., Somerville R.~S., Kravtsov A.~V., Klypin
  A.~A., Primack J.~R., Dekel A., 2001, MNRAS, 321, 559

\bibitem[{Calzetti {et~al}\mbox{.}(2000)Calzetti, Armus, Bohlin, Kinney,
  Koornneef, \& Storchi‐Bergmann}]{Calzetti2000}
Calzetti D., Armus L., Bohlin R.~C., Kinney A.~L., Koornneef J.,
  Storchi‐Bergmann T., 2000, ApJ, 533, 682

\bibitem[{Conselice \& Arnold(2009)}]{Conselice2009}
Conselice C.~J., Arnold J., 2009, MNRAS, 397, 208

\bibitem[{Conselice, Bershady \& Jangren(2000)Conselice, Bershady, \&
  Jangren}]{Conselice2000}
Conselice C.~J., Bershady M.~A., Jangren A., 2000, ApJ, 529, 886

\bibitem[{Curtis-Lake {et~al}\mbox{.}(2013)Curtis-Lake, McLure, Dunlop,
  Schenker, Rogers, Targett, Cirasuolo, Almaini, Ashby, Bradshaw, Finkelstein,
  Dickinson, Ellis, Faber, Fazio, Ferguson, Fontana, Grogin, Hartley, Kocevski,
  Koekemoer, Lai, Robertson, Vanzella, \& Willner}]{Curtis-Lake2013}
Curtis-Lake E. {et~al.}, 2013, MNRAS, 429, 302

\bibitem[{Danovich {et~al}\mbox{.}(2015)Danovich, Dekel, Hahn, Ceverino, \&
  Primack}]{Danovich2014}
Danovich M., Dekel A., Hahn O., Ceverino D., Primack J., 2015, MNRAS, 449, 2087

\bibitem[{Danovich {et~al}\mbox{.}(2012)Danovich, Dekel, Hahn, \&
  Teyssier}]{Danovich2012}
Danovich M., Dekel A., Hahn O., Teyssier R., 2012, MNRAS, 422, 1732

\bibitem[{Ellis {et~al}\mbox{.}(2013)Ellis, McLure, Dunlop, Robertson, Ono,
  Schenker, Koekemoer, Bowler, Ouchi, Rogers, Curtis-Lake, Schneider, Charlot,
  Stark, Furlanetto, \& Cirasuolo}]{Ellis2013}
Ellis R.~S. {et~al.}, 2013, ApJ, 763, L7

\bibitem[{Elmegreen \& Elmegreen(2005)}]{Elmegreen2005}
Elmegreen B.~G., Elmegreen D.~M., 2005, ApJ, 627, 632

\bibitem[{Fall \& Efstathiou(1980)}]{FallS.M.1980}
Fall S.~M., Efstathiou G., 1980, MNRAS, 193, 189

\bibitem[{Fathi {et~al}\mbox{.}(2012)Fathi, Gatchell, Hatziminaoglou, \&
  Epinat}]{Fathi2012}
Fathi K., Gatchell M., Hatziminaoglou E., Epinat B., 2012, MNRAS: Letters, 423,
  L112

\bibitem[{Ferguson {et~al}\mbox{.}(2004)Ferguson, Dickinson, Giavalisco,
  Kretchmer, Ravindranath, Idzi, Taylor, Conselice, Fall, Gardner, Livio,
  Madau, Moustakas, Papovich, Somerville, Spinrad, \& Stern}]{Ferguson2004}
Ferguson H.~C. {et~al.}, 2004, ApJ, 600, L107

\bibitem[{Giavalisco {et~al}\mbox{.}(2004)Giavalisco, Ferguson, Koekemoer,
  Dickinson, Alexander, Bauer, Bergeron, Biagetti, Brandt, Casertano, Cesarsky,
  Chatzichristou, Conselice, Cristiani, {Da Costa}, Dahlen, de~Mello,
  Eisenhardt, Erben, Fall, Fassnacht, Fosbury, Fruchter, Gardner, Grogin, Hook,
  Hornschemeier, Idzi, Jogee, Kretchmer, Laidler, Lee, Livio, Lucas, Madau,
  Mobasher, Moustakas, Nonino, Padovani, Papovich, Park, Ravindranath, Renzini,
  Richardson, Riess, Rosati, Schirmer, Schreier, Somerville, Spinrad, Stern,
  Stiavelli, Strolger, Urry, Vandame, Williams, \& Wolf}]{Giavalisco2004}
Giavalisco M. {et~al.}, 2004, ApJ, 600, L93

\bibitem[{Grazian {et~al}\mbox{.}(2012)Grazian, Castellano, Fontana,
  Pentericci, Dunlop, McLure, Koekemoer, Dickinson, Faber, Ferguson, Galametz,
  Giavalisco, Grogin, Hathi, Kocevski, Lai, Newman, \&
  Vanzella}]{GrazianA.2012}
Grazian A. {et~al.}, 2012, Astronomy {\&} Astrophysics, 547, 51

\bibitem[{Grogin {et~al}\mbox{.}(2011)Grogin, Kocevski, Faber, Ferguson,
  Koekemoer, Riess, Acquaviva, Alexander, Almaini, Ashby, Barden, Bell,
  Bournaud, Brown, Caputi, Casertano, Cassata, Challis, Chary, Cheung,
  Cirasuolo, Conselice, {Roshan Cooray}, Croton, Daddi, Dahlen, Dav{\'{e}},
  de~Mello, Dekel, Dickinson, Dolch, Donley, Dunlop, Dutton, Elbaz, Fazio,
  Filippenko, Finkelstein, Fontana, Gardner, Garnavich, Gawiser, Giavalisco,
  Grazian, Guo, Hathi, H{\"{a}}ussler, Hopkins, Huang, Huang, Jha, Kartaltepe,
  Kirshner, Koo, Lai, Lee, Li, Lotz, Lucas, Madau, McCarthy, McGrath, McIntosh,
  McLure, Mobasher, Moustakas, Mozena, Nandra, Newman, Niemi, Noeske, Papovich,
  Pentericci, Pope, Primack, Rajan, Ravindranath, Reddy, Renzini, Rix, Robaina,
  Rodney, Rosario, Rosati, Salimbeni, Scarlata, Siana, Simard, Smidt,
  Somerville, Spinrad, Straughn, Strolger, Telford, Teplitz, Trump, van~der
  Wel, Villforth, Wechsler, Weiner, Wiklind, Wild, Wilson, Wuyts, Yan, \&
  Yun}]{Grogin2011}
Grogin N.~A. {et~al.}, 2011, ApJS, 197, 39

\bibitem[{Hathi {et~al}\mbox{.}(2008)Hathi, Jansen, Windhorst, Cohen, Keel,
  Corbin, \& {Ryan Jr.}}]{Hathi2008a}
Hathi N.~P., Jansen R.~A., Windhorst R.~A., Cohen S.~H., Keel W.~C., Corbin
  M.~R., {Ryan Jr.} R.~E., 2008, AJ, 135, 156

\bibitem[{Huang {et~al}\mbox{.}(2013)Huang, Ferguson, Ravindranath, \&
  Su}]{Huang2013}
Huang K.-H., Ferguson H.~C., Ravindranath S., Su J., 2013, ApJ, 765, 68

\bibitem[{Ilbert {et~al}\mbox{.}(2009)Ilbert, Capak, Salvato, Aussel,
  McCracken, Sanders, Scoville, Kartaltepe, Arnouts, Floc'h, Mobasher,
  Taniguchi, Lamareille, Leauthaud, Sasaki, Thompson, Zamojski, Zamorani,
  Bardelli, Bolzonella, Bongiorno, Brusa, Caputi, Carollo, Contini, Cook,
  Coppa, Cucciati, {de La Torre}, de~Ravel, Franzetti, Garilli, Hasinger,
  Iovino, Kampczyk, Kneib, Knobel, Kovac, {Le Borgne}, {Le Brun}, F{\`{e}}vre,
  Lilly, Looper, Maier, Mainieri, Mellier, Mignoli, Murayama, Pell{\`{o}},
  Peng, P{\'{e}}rez-Montero, Renzini, Ricciardelli, Schiminovich, Scodeggio,
  Shioya, Silverman, Surace, Tanaka, Tasca, Tresse, Vergani, \&
  Zucca}]{Ilbert2009}
Ilbert O. {et~al.}, 2009, ApJ, 690, 1236

\bibitem[{Jiang {et~al}\mbox{.}(2013)Jiang, Egami, Fan, Windhorst, Cohen, Dave,
  Finlator, Kashikawa, Mechtley, Ouchi, \& Shimasaku}]{Jiang2013a}
Jiang L. {et~al.}, 2013, ApJ, 773, 153

\bibitem[{Kawamata {et~al}\mbox{.}(2015)Kawamata, Ishigaki, Shimasaku, Oguri,
  \& Ouchi}]{Kawamata2014}
Kawamata R., Ishigaki M., Shimasaku K., Oguri M., Ouchi M., 2015, ApJ, 804, 103

\bibitem[{Koekemoer {et~al}\mbox{.}(2013)Koekemoer, Ellis, McLure, Dunlop,
  Robertson, Ono, Schenker, Ouchi, Bowler, Rogers, Curtis-Lake, Schneider,
  Charlot, Stark, Furlanetto, Cirasuolo, Wild, \& Targett}]{Koekemoer2013}
Koekemoer A.~M. {et~al.}, 2013, ApJS, 209, 3

\bibitem[{Koekemoer {et~al}\mbox{.}(2011)Koekemoer, Faber, Ferguson, Grogin,
  Kocevski, Koo, Lai, Lotz, Lucas, McGrath, Ogaz, Rajan, Riess, Rodney,
  Strolger, Casertano, Dahlen, Dickinson, Dolch, Fontana, Giavalisco, Grazian,
  Guo, Hathi, Huang, van~der Wel, Yan, Acquaviva, Almaini, Ashby, Barden, Bell,
  Bournaud, Brown, Caputi, Cassata, Challis, Chary, Cheung, Cirasuolo,
  Conselice, {Roshan Cooray}, Croton, Daddi, Dav{\'{e}}, de~Mello, de~Ravel,
  Dekel, Donley, Dunlop, Dutton, Elbaz, Fazio, Filippenko, Finkelstein, Frazer,
  Gardner, Garnavich, Gawiser, Gruetzbauch, Hartley, H{\"{a}}ussler,
  Herrington, Hopkins, Huang, Jha, Johnson, Kartaltepe, Khostovan, Kirshner,
  Lani, Lee, Li, Madau, McCarthy, McIntosh, McLure, McPartland, Mobasher,
  Moreira, Mortlock, Moustakas, Mozena, Nandra, Newman, Nielsen, Niemi, Noeske,
  Papovich, Pentericci, Pope, Primack, Ravindranath, Reddy, Renzini, Rix,
  Robaina, Rosario, Rosati, Salimbeni, Scarlata, Siana, Simard, Smidt, Snyder,
  Somerville, Spinrad, Straughn, Telford, Teplitz, Trump, Vargas, Villforth,
  Wagner, Wandro, Wechsler, Weiner, Wiklind, Wild, Wilson, Wuyts, \&
  Yun}]{Koekemoer2011}
Koekemoer A.~M. {et~al.}, 2011, ApJS, 197, 36

\bibitem[{Kron(1980)}]{Kron1980}
Kron R.~G., 1980, ApJS, 43, 305

\bibitem[{Law {et~al}\mbox{.}(2009)Law, Steidel, Erb, Larkin, Pettini, Shapley,
  \& Wright}]{Law2009}
Law D.~R., Steidel C.~C., Erb D.~K., Larkin J.~E., Pettini M., Shapley A.~E.,
  Wright S.~A., 2009, ApJ, 697, 2057

\bibitem[{Livermore {et~al}\mbox{.}(2015)Livermore, Jones, Richard, Bower,
  Swinbank, Yuan, Edge, Ellis, Kewley, Smail, Coppin, \&
  Ebeling}]{Livermore2015}
Livermore R.~C. {et~al.}, 2015, MNRAS, 450, 1812

\bibitem[{Lorenzoni {et~al}\mbox{.}(2012)Lorenzoni, Bunker, Wilkins, Caruana,
  Stanway, \& Jarvis}]{Lorenzoni2012}
Lorenzoni S., Bunker A.~J., Wilkins S.~M., Caruana J., Stanway E.~R., Jarvis
  M.~J., 2012, MNRAS, 429, 150

\bibitem[{Madau(1995)}]{Madau1995}
Madau P., 1995, ApJ, 441, 18

\bibitem[{McLure {et~al}\mbox{.}(2009)McLure, Cirasuolo, Dunlop, Foucaud, \&
  Almaini}]{McLure2009}
McLure R.~J., Cirasuolo M., Dunlop J.~S., Foucaud S., Almaini O., 2009, MNRAS,
  395, 2196

\bibitem[{McLure {et~al}\mbox{.}(2013)McLure, Dunlop, Bowler, Curtis-Lake,
  Schenker, Ellis, Robertson, Koekemoer, Rogers, Ono, Ouchi, Charlot, Wild,
  Stark, Cirasuolo, \& Targett}]{McLure2013}
McLure R.~J. {et~al.}, 2013, MNRAS, 432, 2696

\bibitem[{Mo, Mao \& White(1998)Mo, Mao, \& White}]{Mo1998}
Mo H.~J., Mao S., White S. D.~M., 1998, MNRAS, 295, 319

\bibitem[{Oesch {et~al}\mbox{.}(2010{\natexlab{a}})Oesch, Bouwens, Carollo,
  Illingworth, Trenti, Stiavelli, Magee, Labb{\'{e}}, \& Franx}]{Oesch2010a}
Oesch P.~A. {et~al.}, 2010{\natexlab{a}}, ApJ, 709, L21

\bibitem[{Oesch {et~al}\mbox{.}(2010{\natexlab{b}})Oesch, Bouwens, Illingworth,
  Carollo, Franx, Labb{\'{e}}, Magee, Stiavelli, Trenti, \& van
  Dokkum}]{Oesch2010}
Oesch P.~A. {et~al.}, 2010{\natexlab{b}}, ApJ, 709, L16

\bibitem[{Ono {et~al}\mbox{.}(2013)Ono, Ouchi, Curtis-Lake, Schenker, Ellis,
  McLure, Dunlop, Robertson, Koekemoer, Bowler, Rogers, Schneider, Charlot,
  Stark, Shimasaku, Furlanetto, \& Cirasuolo}]{Ono2013}
Ono Y. {et~al.}, 2013, ApJ, 777, 155

\bibitem[{Padilla {et~al}\mbox{.}(2014)Padilla, Salazar, Contreras, Cora, \&
  Ruiz}]{Padilla2014}
Padilla N., Salazar S., Contreras S., Cora S., Ruiz A., 2014, MNRAS, 443, 2801

\bibitem[{Polletta {et~al}\mbox{.}(2007)Polletta, Tajer, Maraschi, Trinchieri,
  Lonsdale, Chiappetti, Andreon, Pierre, {Le Fevre}, Zamorani, Maccagni,
  Garcet, Surdej, Franceschini, Alloin, Shupe, Surace, Fang, Rowan‐Robinson,
  Smith, \& Tresse}]{Polletta2007}
Polletta M. {et~al.}, 2007, ApJ, 663, 81

\bibitem[{Sales {et~al}\mbox{.}(2012)Sales, Navarro, Theuns, Schaye, White,
  Frenk, Crain, \& {Dalla Vecchia}}]{Sales2012}
Sales L.~V., Navarro J.~F., Theuns T., Schaye J., White S. D.~M., Frenk C.~S.,
  Crain R.~A., {Dalla Vecchia} C., 2012, MNRAS, 423, 1544

\bibitem[{Schenker {et~al}\mbox{.}(2013)Schenker, Robertson, Ellis, Ono,
  McLure, Dunlop, Koekemoer, Bowler, Ouchi, Curtis-Lake, Rogers, Schneider,
  Charlot, Stark, Furlanetto, \& Cirasuolo}]{Schenker2013}
Schenker M.~A. {et~al.}, 2013, ApJ, 768, 196

\bibitem[{Scoville {et~al}\mbox{.}(2007)Scoville, Aussel, Brusa, Capak,
  Carollo, Elvis, Giavalisco, Guzzo, Hasinger, Impey, Kneib, LeFevre, Lilly,
  Mobasher, Renzini, Rich, Sanders, Schinnerer, Schminovich, Shopbell,
  Taniguchi, \& Tyson}]{Scoville2007}
Scoville N. {et~al.}, 2007, ApJS, 172, 1

\bibitem[{Shibuya, Ouchi \& Harikane(2015)Shibuya, Ouchi, \&
  Harikane}]{Shibuya2015}
Shibuya T., Ouchi M., Harikane Y., 2015, ApJS, 219, 15

\bibitem[{Stark {et~al}\mbox{.}(2009)Stark, Ellis, Bunker, Bundy, Targett,
  Benson, \& Lacy}]{Stark2009}
Stark D.~P., Ellis R.~S., Bunker A., Bundy K., Targett T., Benson A., Lacy M.,
  2009, ApJ, 697, 1493

\bibitem[{Stark {et~al}\mbox{.}(2013)Stark, Schenker, Ellis, Robertson, McLure,
  \& Dunlop}]{Stark2013}
Stark D.~P., Schenker M.~A., Ellis R.~S., Robertson B., McLure R., Dunlop J.,
  2013, ApJ, 763, 129S

\bibitem[{Windhorst {et~al}\mbox{.}(2011)Windhorst, Cohen, Hathi, McCarthy,
  Ryan, Yan, Baldry, Driver, Frogel, Hill, Kelvin, Koekemoer, Mechtley,
  O'Connell, Robotham, Rutkowski, Seibert, Straughn, Tuffs, Balick, Bond,
  Bushouse, Calzetti, Crockett, Disney, Dopita, Hall, Holtzman, Kaviraj,
  Kimble, MacKenty, Mutchler, Paresce, Saha, Silk, Trauger, Walker, Whitmore,
  \& Young}]{Windhorst2011}
Windhorst R.~A. {et~al.}, 2011, ApJS, 193, 27

\end{thebibliography}

\appendix
\section{Impact of choice of S\'ersic index in simulations}
\label{Appendix}

\begin{figure}
  \centering
  \includegraphics[width=3.4in]{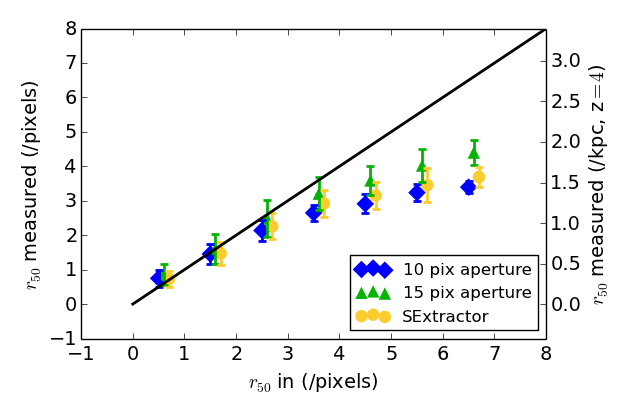}
  \caption{Measured half-light radius vs. true half-light radius for all profiles with $m_{tot}<27$ for single S\'ersic profiles with $n=2$.  The measured sizes are binned according to input size with $\Delta r_{50}=1$, and the medians and standard deviations are plotted as the points and error bars respectively.  Three different measurements are plotted as indicated in the legend.}
  \label{fig:diagnosticPlot_n2}
\end{figure}

\begin{figure}
  \centering
  \subfigure[Mean size estimates]{\includegraphics[width=3.4in]{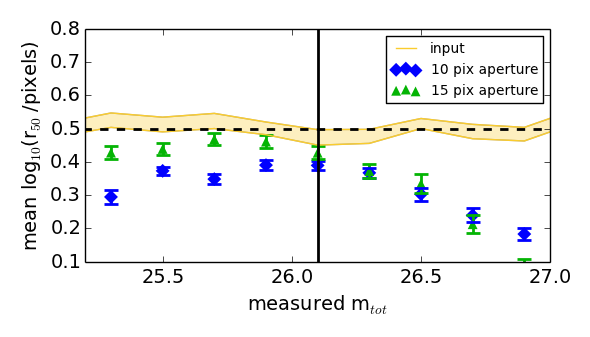}}
  \subfigure[Modal size estimates]{\includegraphics[width=3.4in]{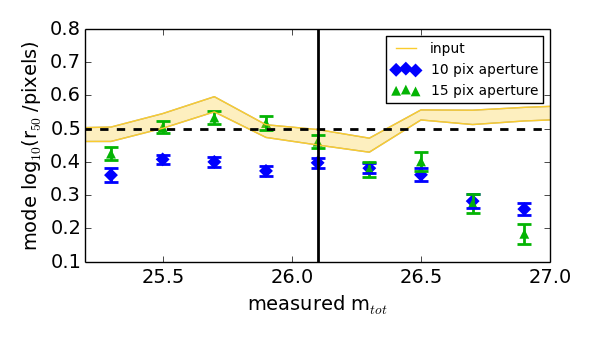}}
  \caption{Typical galaxy size measurements as shown in Fig. 4 but for single S\'ersic profiles with $n=2$.}
  \label{fig:typicalSizeBias_n2}
\end{figure}

Here we show the results of simulation sets I and II using single S\'ersic profiles with $n=2$.  In Fig. \ref{fig:diagnosticPlot_n2} we display the measured vs. input sizes (as for Fig. \ref{fig:diagnosticPlot}, upper left panel) and in Fig.  \ref{fig:typicalSizeBias_n2} the typical size measurements are plotted (as for Fig. \ref{fig:typicalSizeBias}).

We see from Fig. \ref{fig:diagnosticPlot_n2} that all measurement techniques start to systematically underestimate the half-light radii at smaller sizes than for $n=1$ profiles.  The sizes for input half-light radii larger than $\sim2$ pixels ($\sim0.9$kpc at $z\sim4$) are underestimated for both the 10 pixel aperture and when using {\scshape SExtractor}. The 15 pixel aperture becomes biased at very slightly larger sizes.  Fig. \ref{fig:typicalSizeBias_n2} shows that the typical size estimates using the mean are biased to small sizes for both total flux apertures. The modal sizes are also underestimated with the 10 pixel aperture but not for the 15 pixel aperture as the measured sizes are better reproduced with this sized aperture at the peak of the simulated size distribution ( $\sim1.3$ kpc).

These results bring up the question of whether our results at $z\sim4$ might be underestimated if the intrinsic profiles are closer to $n=2$ profiles than $n=1$.  However, if the intrinsic size distribution is peaked at $\sim1.3$ kpc, we would expect to measure a higher value than we do at $z\sim4$ (we measure $\sim0.9$ kpc from the sample and $\sim1.08$ kpc from the simulated distribution).  At the actual measured typical size ($\sim0.9$ kpc), the measurements shouldn't be underestimated according to Fig. \ref{fig:diagnosticPlot_n2}. In fact, the size used for these simulations ($\sim1.3$ kpc) is taken from previous measurements that measure the mean of the distribution in real space and reproducing this measurement using the 15 pixel aperture with our sample agrees well (see Fig. \ref{fig:size_redshift_bright}).  This suggests that the actual peak of the distribution is, indeed smaller than previous size estimates.  We can also use the fact that the modal estimate from the 15 pix aperture is found to be un-biased and compare the estimate with this larger flux aperture to that derived with the 10 pixel aperture.  We find it is marginally larger, but not significantly so, and not enough to alter the derived evolution if this modal estimate is used rather than that for the 10 pixel aperture.

\section{Star formation clump simulations}
\label{Appendix:clumps}

To give physical context to our measurements of no evolution in the disturbed fraction of galaxies we investigate what physical properties clumps of star formation require to produce a disturbed morphology with our measurement technique.  The main results are referred to and summarised in Section \ref{section:discussionClumpy}.

For these simulations we add single clumps to single S\'ersic profiles with the range of properties outlined in Table \ref{table:simulations} for simulation set III.  They have a uniform distribution of S\'ersic profiles in the range $0.5<n<4.5$, uniform distribution of sizes in the range $0.5 < r_{50} ~\textrm{(/pix)}<10$ , uniform distribution of axis ratios between $0.2 < ar < 1$ and uniform distribution of total magnitudes between $23 < $\MUV $<31$.  Single clumps are added to each object with magnitudes in the range $26<m_c<29$.  We look at both un-resolved clumps and clumps with Gaussian profiles (without large wings) with sizes distributed between $0.1 < r_c/r_e<2$ .  The figures presented here are for unresolved clumps only.  These simulations are performed on profiles inserted into the $Y_{105}$ GOODS-S deep image (see Section \ref{section:discussionClumpy}).

From these simulations we find that the main variables that affect the asymmetry measurement are the distance of the clump from the centre of the galaxy (must be $>3$ pixels), the difference in brightness between the clump and the profile and the brightness of the clump itself (the clumps must have a minimum surface brightness before they can alter the asymmetry measurement).

\begin{figure}
  \centering
  \subfigure[]{\includegraphics[width=3.4in]{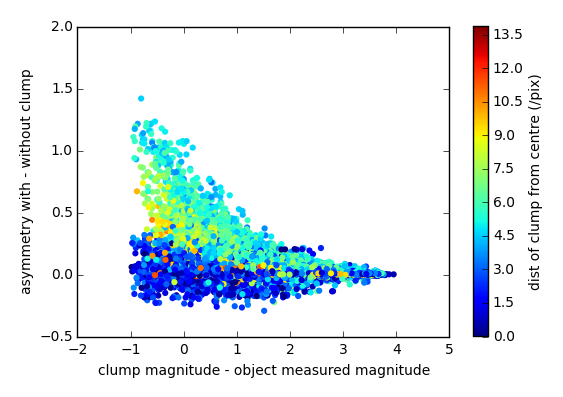}}
  \subfigure[]{\includegraphics[width=3.4in, trim= 2cm 0cm 0cm 0cm,clip]{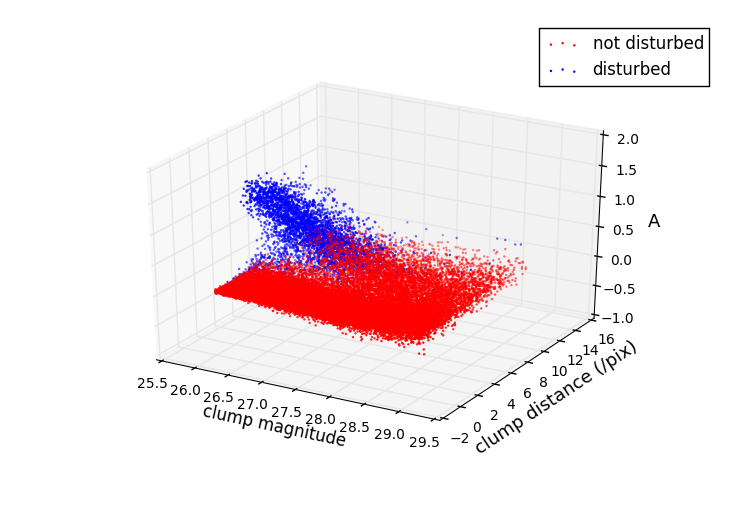}}
  \caption{Showing the change in asymmetry as a function of clump parameters.  \textbf{Panel (a)} shows the change in measured asymmetry as a function of the difference between clump and object magnitude.  The points are colour-coded according to the distance of the clump from the centre of the object. \textbf{Panel (b)} shows the measured asymmetry as a function of both clump magnitude and distance of clump from centre of object with the colour of the point showing whether or not the object is then measured to be disturbed.}
  \label{fig:clumpAppendix1}
  \end{figure}

Fig. \ref{fig:clumpAppendix1} shows the results for un-resolved clumps of star formation.  It shows the change in clump asymmetry as a function of the difference in clump and object magnitude, colour-coded according to the distance of the clump from the centre of the object.  It shows that clumps must be at least 3 pixels from the centre of the object to allow the clump and object centre to be separated.  Within this distance the clump becomes merged with the centre and the asymmetry measurement will be minimised by re-centering the object.  The clump will also not change the asymmetry if it is too far from the centre of the object because the Asymmetry is only measured within $r_{70}$.  

We find that there is a minimum clump magnitude required, and that 50\% of objects with clumps of 27.25 magnitudes have disturbed morphologies.   The clump is also required have a minimum brightness contrast with respect to the object and the distribution is shown to decrease rapidly for $\Delta m>1$, or $f_c/f_g<0.4$ (panel b).  This is only mildly dependent on the profile of the underlying object, with discier galaxies showing disturbed morphologies for fainter clumps.  We also investigate whether these results are dependent on clump size and find that surface brightness is more important than fraction of the object surface area occupied by the clump, so resolved clumps must have higher total fluxes to produce a disturbed morphology.

\label{lastpage}

\end{document}